\documentclass[pra,twocolumn,aps,showpacs,amsmath,amssymb,superscriptaddress,floatfix]{revtex4-2} 

\usepackage{color}
\usepackage{bm}
\usepackage{dcolumn}
\usepackage{graphicx}
\usepackage{braket}
\usepackage{lipsum}
\usepackage{multirow}

\definecolor{royalblue}{RGB}{4,51,255}
\definecolor{eggplant}{RGB}{153,25,94}

\newcommand{\ms}[1]{{\color{eggplant}[M: #1]}}
\newcommand{\AW}[1]{{\color{red}{#1}}}
\newcommand{\mf}[1]{{\color{orange}{MF: #1}}}
\newcommand{\jp}[1]{{\color{blue}{#1}}}

\usepackage{xcolor}
\usepackage{xparse,xcoffins}
\ExplSyntaxOn
\NewCoffin\imagecoffin
\NewCoffin\labelcoffin
\keys_define:nn { miguel/label }
 {
  label   .tl_set:N = \l_miguel_label_tl,
  labelbox .bool_set:N = \l_miguel_label_box_bool,
  labelbox .default:n = true,
  fontsize .tl_set:N = \l_miguel_label_size_tl,
  fontsize .initial:n = \footnotesize,
  pos .choice:,
  pos/nw .code:n = \tl_set:Nn \l_miguel_label_pos_tl { left,up },
  pos/ne .code:n = \tl_set:Nn \l_miguel_label_pos_tl { right,up },
  pos/sw .code:n = \tl_set:Nn \l_miguel_label_pos_tl { left,down },
  pos/se .code:n = \tl_set:Nn \l_miguel_label_pos_tl { right,down },
  pos/n .code:n = \tl_set:Nn \l_miguel_label_pos_tl { hc,up },
  pos/w .code:n = \tl_set:Nn \l_miguel_label_pos_tl { left,vc },
  pos/s .code:n = \tl_set:Nn \l_miguel_label_pos_tl { hc,down },
  pos/e .code:n = \tl_set:Nn \l_miguel_label_pos_tl { right,vc },
  pos .initial:n = nw,
  unknown .code:n   = \clist_put_right:Nx \l_miguel_label_clist
                       { \l_keys_key_tl = \exp_not:n { #1 } }
 }
\clist_new:N \l_miguel_label_clist
\box_new:N \l_miguel_label_box
\box_new:N \l_miguel_label_image_box
\NewDocumentCommand{\xincludegraphics}{O{}m}
 {
  \group_begin:
  \tl_clear:N \l_miguel_label_tl
  \clist_clear:N \l_miguel_label_clist
  \keys_set:nn { miguel/label } { #1 }
  \tl_if_empty:NTF \l_miguel_label_tl
   {
    \miguel_includegraphics:Vn \l_miguel_label_clist { #2 }
   }
   {
    \SetHorizontalCoffin\imagecoffin
     {
      \miguel_includegraphics:Vn \l_miguel_label_clist { #2 }
     }
    \SetHorizontalCoffin\labelcoffin
     {
      \raisebox{\depth}
       {
        \bool_if:NTF \l_miguel_label_box_bool
         { \fcolorbox{white}{white}{\l_miguel_label_size_tl\l_miguel_label_tl} }
         { \l_miguel_label_size_tl\l_miguel_label_tl }
       }
     }
    \SetVerticalPole\imagecoffin{left}{3pt+\CoffinWidth\labelcoffin/2}
    \SetVerticalPole\imagecoffin{right}{\Width-3pt-\CoffinWidth\labelcoffin/2}
    \SetHorizontalPole\imagecoffin{up}{\Height-3pt-\CoffinHeight\labelcoffin/2}
    \SetHorizontalPole\imagecoffin{down}{3pt+\CoffinHeight\labelcoffin/2}
    \use:x{\JoinCoffins\imagecoffin[\l_miguel_label_pos_tl]\labelcoffin[vc,hc]} 
    \TypesetCoffin\imagecoffin
   }
   \group_end:
 }
\NewDocumentCommand{\setlabel}{m}
 {
  \keys_set:nn { miguel/label } { #1 }
 }
\cs_new_protected:Nn \miguel_includegraphics:nn
 {
  \includegraphics[#1]{#2}
 }
\cs_generate_variant:Nn \miguel_includegraphics:nn { V }
\ExplSyntaxOff

\usepackage[pdftex]{hyperref}
\usepackage{ulem} 
\hypersetup{colorlinks = true, urlcolor = blue, linkcolor = blue, citecolor = blue}

\begin{document}

\title{Disentangling Interacting Systems with Fermionic Gaussian Circuits: Application to Quantum Impurity Models}

\author{Ang-Kun Wu}
\affiliation{Department of Physics and Astronomy, Center for Materials Theory, Rutgers University, Piscataway, New Jersey 08854, USA}
\affiliation{Center for Computational Quantum Physics, Flatiron Institute, New York, New York 10010, USA}
\author{Benedikt Kloss}
\affiliation{Center for Computational Quantum Physics, Flatiron Institute, New York, New York 10010, USA}
\author{Wladislaw Krinitsin}
\affiliation{Center for Computational Quantum Physics, Flatiron Institute, New York, New York 10010, USA}
\author{Matthew T. Fishman}
\affiliation{Center for Computational Quantum Physics, Flatiron Institute, New York, New York 10010, USA}
\author{J. H. Pixley}
\affiliation{Department of Physics and Astronomy, Center for Materials Theory, Rutgers University, Piscataway, New Jersey 08854, USA}
\affiliation{Center for Computational Quantum Physics, Flatiron Institute, New York, New York 10010, USA}
\author{E. M. Stoudenmire}
\affiliation{Center for Computational Quantum Physics, Flatiron Institute, New York, New York 10010, USA}
\date{\today}

\begin{abstract}
Tensor network quantum states are powerful tools for strongly correlated systems, tailored to capture local correlations such as in ground states with entanglement area laws. When applying tensor network states to interacting fermionic systems, a proper choice of the basis or orbitals can reduce the bond dimension of tensors and provide physically relevant orbitals. We introduce such a change of basis with unitary gates obtained from compressing fermionic Gaussian states into quantum circuits corresponding to various tensor networks. These circuits can reduce the ground state entanglement entropy and improve the performance of algorithms such as the density matrix renormalization group.  We study the Anderson impurity model with one and two impurities to show the potential of the method for improving computational efficiency and interpreting impurity physics. Furthermore, fermionic Gaussian circuits can also suppress entanglement during the time evolution out of low-energy state. Last, we consider Gaussian multi-scale entanglement renormalization ansatz (GMERA) circuits which compress fermionic Gaussian states hierarchically. The emergent coarse-grained physical models from these GMERA circuits are studied in terms of their entanglement properties and suitability for performing time evolution. 
\end{abstract}

\maketitle


\section{Introduction}

Tensor network methods,  originating from the density matrix renormalization group (DMRG), 
 provide a powerful set of tools to study many-body quantum systems \cite{PhysRevLett.69.2863,PhysRevB.48.10345,RevModPhys.77.259,hallberg2006new,schollwock2011density}.
The organizing principle behind these methods is entanglement. States with low entanglement can be factored into a network of small tensors, leading to fast and precise algorithms. The size of the indices connecting the tensors, known as the bond dimension, is bounded by the amount of entanglement and directly controls the cost of tensor network calculations.

Entanglement is exquisitely sensitive to the basis in which a state is represented. Non-interacting fermion systems exhibit some of the most highly entangled ground states \cite{Evenbly:2011,Barthel:2006,Wolf:2006,PhysRevLett.119.020601,PhysRevB.104.214306,PhysRevB.103.L241118,kiefer2020bounds,ChinPhysLett.40.057102} yet become completely unentangled when transformed into the momentum basis (assuming periodic boundary conditions). 
More general low-energy states of local Hamiltonians typically obey the \emph{area law} (up to logarithmic corrections), meaning the entanglement of subsystem of volume $l^d$ in dimension $d$ scales as its boundary size $\sim l^{d-1}$. Though the area law explains why
low-energy states in a real-space basis have limited entanglement, it does not imply that real space is the basis with the \emph{lowest} entanglement.

The insight that another basis besides real space could yield more efficient tensor network calculations has been put into practice for some time. A notable example is the work by Krumnow, Eisert, and Legeza on computing basis transformations \cite{Krumnow:2016,Krumnow:2019,Krumnow:2021}. Their approach  optimizes local two-site unitary gates to minimize the entanglement across a given bipartition of the system, often resulting in significantly lower entanglement throughout the system, for both one- and two-dimensional lattices and dynamical systems.

In a separate advance, Fishman and White \cite{Fishman:2015} showed how the problem of transforming Gaussian fermionic states into tensor networks could be solved by finding quantum circuits---fermionic Gaussian circuits---which provide a basis where the Gaussian state becomes unentangled. Their approach has many appealing aspects, such as a fast, direct method for computing circuits that preserve the locality of the Hamiltonian. These many-body circuits are equivalent to canonical transformations of the single-particle basis and map straightforwardly into known classes of tensor networks, like matrix product states (MPS) \cite{PhysRevLett.69.2863,PhysRevB.48.10345,PhysRevB.55.2164}, tree tensor networks (TTN) \cite{PhysRevA.74.022320}, and multiscale entanglement renormalization ansatz (MERA) networks \cite{PhysRevLett.99.220405,PhysRevLett.101.110501,PhysRevB.81.235102}. 
A notable aspect of the method is that it takes advantage of the large eigenvalue degeneracy of the correlation matrix to constrain the computed orbitals to be local. This method also gives insights into how single-particle basis transformations are related to many-body entanglement.

The relevance of the Gaussian circuits found by Fishman and White for interacting, non-Gaussian systems has been unclear. In this work, we synthesize the above ideas, utilizing basis transformations obtained from noninteracting systems to reduce the entanglement of interacting systems. 
We find that Gaussian circuit transformations can both speed up tensor network calculations and yield physical insights.
To demonstrate our approach, we consider the quantum impurity models in one dimension \cite{PhysRev.124.41}, whose physical properties and entanglement structure are known and provide a rich testbed for our methods.

After demonstrating that Gaussian circuits can uncover a basis in which the ground state has low entanglement, we show how certain physical properties, such as Kondo screening and the RKKY effect,
 become more transparent in the new basis. The same approach can also reduce the entanglement of a system undergoing quench dynamics. 

Beyond simpler Gaussian circuit transformations based on matrix product states, 
we also consider multiscale circuits known as Gaussian MERA (GMERA). In addition to reducing entanglement, 
GMERA circuits provide a physical interpretation of transformed degrees of freedom. The basis introduced by a conventional GMERA is analogous to a ``star geometry'' bath used in impurity solvers \cite{bauernfeind2019comparison,PhysRevB.90.235131}
while a modified boundary GMERA \cite{PhysRevB.82.161107,evenbly2014algorithms,PhysRevB.91.205119} is analogous to a Wilson ``chain geometry'' bath \cite{RevModPhys.47.773,RevModPhys.80.395}. 
The GMERA approach lets us identify an important subset of sites in the transformed basis, and we
show that reducing the bath to consist of only these sites is sufficient to obtain dynamical properties.

We begin by reviewing the technology of fermionic Gaussian circuits in Sec.~\ref{sec:fgg}. After introducing the impurity models 
in Sec.~\ref{sec:model}, we explore how Gaussian MPS (GMPS) circuits can drastically reduce their entanglement in Sec.~\ref{sec:disentangler}, resulting in much faster DMRG solutions without ever passing through a more highly entangled basis. To discuss the physical insights in these fermionic Gaussian circuits, we show how the Gaussian circuit transformations reveal Kondo screening physics in Sec.~\ref{sec:screening}, and we discuss two-impurity models where inter-impurity interactions 
arise due to the RKKY effect and how it emerges in the transformed basis in \ref{sec:twoimpurity}.
In Sec.~\ref{sec:timeEvolution}, we demonstrate that the disentangling can succeed in systems undergoing dynamics, too. 
Finally, Sec.~\ref{sec:GMERA} explores two types of GMERA circuits, the physical interpretation of the noninteracting bath following each GMERA transformation, and the possibility of simulating the impurity model in a smaller coarse-grained space. 

\section{Review: Fermionic Gaussian Tensor Network Circuits}
\label{sec:fgg}

Before turning to interacting systems, we briefly review the technology of compressing fermionic Gaussian states, following Ref.~\cite{Fishman:2015}, through which \emph{fermionic Gaussian circuits}  \cite{Fishman:2015,surace2022fermionic} (also known as \emph{matchgate circuits} \cite{valiant2001quantum,jozsa2008matchgates}) are defined. 
Different layouts and patterns of these circuits produce various Gaussian tensor networks, such as the GMPS and the GMERA circuits.

Given a noninteracting system with a Hamiltonian
\begin{align}\label{eq:H0}
    H_0 = \sum_{ij} h_{ij} c^\dagger_i c_j,
\end{align}
and $c_i^\dagger$ the spinless fermionic creation operator at site $i$,
one can straightforwardly construct its ground state by diagonalizing the matrix $h = U D U^\dagger$ and filling the orbitals (columns of $U$) having the lowest energies (eigenvalues of $h$). The ground state can be written as
\begin{align}
    \ket{\psi_0} =\prod_{k=1}^{N_F} c_k^\dagger \ket{\Omega}, 
\end{align}
where $N_F$ is the number of fermions and $\ket{\Omega}$ is the vacuum.

The construction of fermionic Gaussian circuits is based on the correlation matrix
\begin{align}
    \Lambda_{ij} = \bra{\psi_0}c_{i}^\dagger c_{j}\ket{\psi_0}=\sum_{k=1}^{N_F} U^*_{ik} U_{jk},
\end{align}
where the goal is to approximately diagonalize $\Lambda$ with a circuit that also preserves locality.
In a basis where $\Lambda$ is diagonal, the state will have zero entanglement \cite{Vidal:2003,Fishman:2015}.

An important fact about the correlation matrix is that, for a pure and number conserving fermionic state, its eigenvalues are all exactly $0$ or $1$, corresponding to empty or filled orbitals. The presence of many orbitals with degenerate eigenvalues provides room to balance locality constraints with finding good approximations to eigenvectors.
For one-dimensional (1D) states obeying the entanglement area law, such as ground states of gapped and local Hamiltonians, there exists a contiguous subblock $\mathcal{B}$ of size $B$ that is independent of system size $N$ in which one of the eigenvalues will be close to 0 or 1. For ground states of gapless 1D systems, the block size needed to find an eigenvalue close to 0 or 1 only grows modestly with system size as $\log{N}$ \cite{Fishman:2015}.

Here we give a brief summary of the deterministic algorithm for diagonalizing a correlation matrix with local rotations. Local rotations of the form:
\begin{align}
    R(\theta) = \begin{pmatrix}
    \cos\theta & -\sin\theta \\
    \sin\theta & \cos\theta
    \end{pmatrix} \label{eq:Givens}
\end{align}
are used to transform $2 \times 2$ blocks of $\mathcal{B}$, much like in the $QR$ factorization of a matrix based on Givens rotations. The rotations $R(\theta)$ are chosen to isolate an \emph{inactive orbital} or eigenvector of $\mathcal{B}$ with eigenvalue close to $0$ or $1$ in the upper-left corner of the block $\mathcal{B}$ of $\Lambda$. The rotation angle $\theta$ is determined numerically from an eigenvector, for example if the eigenvector is $v=(v_1,v_2)^T$, the angle $\theta=-\arctan\frac{v2}{v1}$ rotates the vector to $v'=(1,0)^T.$ Higher dimensional eigenvectors can be rotated iteratively \cite{Fishman:2015}.
In the resulting basis, the first site of the transformed block will correspond to the inactive orbital which will be approximately disentangled 
from the rest of the system.

For each subblock $\mathcal{B}$, one can obtain a minimal set of local $B-1$ rotation gates which act together as 
$R_\mathcal{B}$
that isolate an inactive orbital of $\mathcal{B}$ through $B-1$ rotations.  After discarding the row and column corresponding to the
disentangled (inactive) site, one can repeat the process on a new block, expanding the block as needed to find the next inactive orbital by including adjacent rows and columns of the correlation matrix. Once all blocks have 
had their most inactive orbital disentangled, the entire correlation matrix will be approximately diagonalized by a sequence of gates
\begin{equation}
    R\equiv R_{\mathcal{B}_1}R_{\mathcal{B}_2}\cdots R_{\mathcal{B}_{N-1}}, 
\end{equation}
as described in detail in Refs. \cite{Fishman:2015,surace2022fermionic}. At the single-particle level, the matrix $R$ defines a transformation of
the $c_i$ operators into new operators 
\begin{equation} \label{eq:transformation}
    f_r = \sum_i R_{ri} c_i,
\end{equation}
which annihilate fermions in the transformed basis indexed by $r$. 
These orbitals generally differ from the single-particle eigenstates, in that they are local and do not diagonalize the quadratic Hamiltonian.

\begin{figure}[t!]
\begin{center}
\includegraphics[width = 0.4\textwidth]{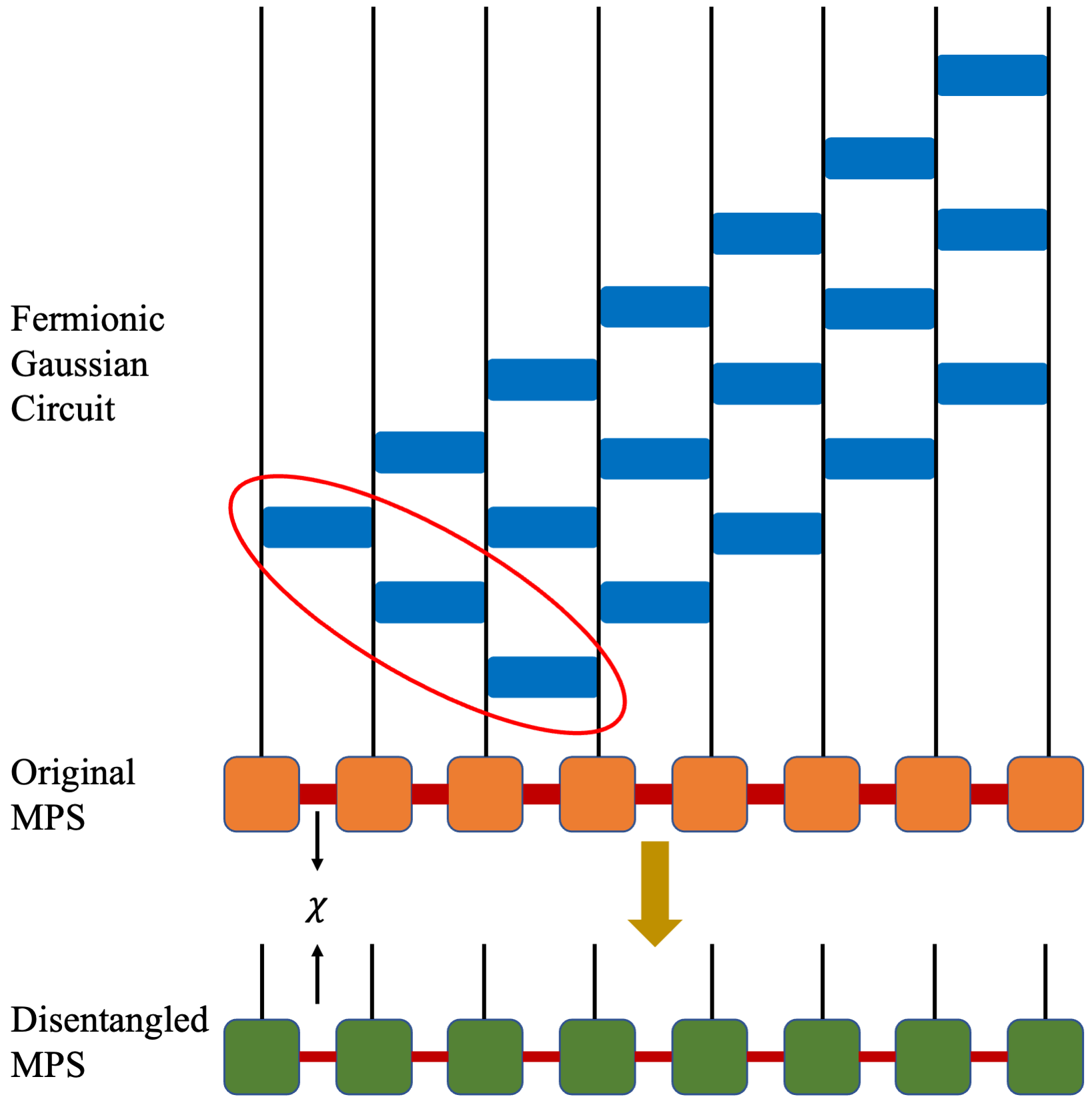}
\caption{\label{fig:Demo}  Demonstration of a fermionic Gaussian circuit computed from a fermionic Gaussian state being applied to an MPS. The MPS ground state in the original basis is composed of smaller tensors (orange squares) with physical dimensions (black lines) and connected by bonds (red edges). After applying the conjugated and reversed fermionic Gaussian gates $\hat{R}(\theta_i)^\dagger$ (blue rectangles) with fixed block size $B=4$, the MPS is transformed into a new basis (green squares), hopefully with a smaller bond dimensions $\chi$. The $B-1$ gates acting on a single block of size $B=4$, denoted in the text by $\hat{R}_{\mathcal{B}}$, are marked by the red ellipse. In the case when the original MPS is a fermionic Gaussian state and the fermionic Gaussian circuit is constructed from its correlation matrix, the disentangled state will have a bond dimension of $\chi=1$, approximately disentangled up to controllable errors in the circuit construction.
}
\end{center}
\end{figure}

Once the set of single-particle transformations $R$ have been found, these transformations can be 
``lifted'' or promoted into many-body quantum gates forming a quantum circuit. We show an example of an (inverse) GMPS circuit acting on
a many-body MPS in Fig.~\ref{fig:Demo}. Specifically, an individual matrix $R(\theta)$ acting on the $(j,j+1)$ block of the correlation matrix can
be reinterpreted as a unitary gate $\hat{R}_{j,j+1}(\theta)$ acting on sites $j,j+1$ and having the following form
\begin{equation} \label{eq:matchgate}
    \hat{R}(\theta) = \begin{pmatrix}
    1 & 0 & 0 & 0 \\
    0 & \cos\theta & -\sin\theta & 0\\
    0 & \sin\theta & \cos\theta & 0\\
    0 & 0 & 0 & 1
    \end{pmatrix}
\end{equation}
The $4$ by $4$ matrix above acts in the basis $\{\ket{\Omega}, c^\dagger_j \ket{\Omega}, c^\dagger_{j+1} \ket{\Omega}, c^\dagger_j c^\dagger_{j+1}\ket{\Omega}\}$ of the $j$ and $j+1$ sites.
These $\hat{R}_{j,j+1}(\theta)$ are the blue tensors shown in the circuit at the top of Fig.~\ref{fig:Demo}.
The red ellipse in the figure highlights the tensors acting within the first block $\mathcal{B}_1=\{1,2,3,4\}$, assuming a block size of $B=4$.

Fermionic Gaussian circuits obtained as described above are called Gaussian matrix product state (GMPS) circuits. A slight modification of the procedure leads to Gaussian MERA (GMERA) circuits discussed in Sec.~\ref{sec:GMERA}. Once the circuits are obtained, they can be used to efficiently transform and disentangle many-body states in the MPS representation as shown in Fig.~\ref{fig:Demo}. 

The techniques above can be straightforwardly extended to apply to fermions with spin or to systems not preserving fermion number \cite{Thoenniss}, though we will continue to use the spinless fermion formalism for reasons explained in the next section.

\section{Impurity Model Systems}
\label{sec:model}

To explore the ability of Gaussian circuits to disentangle interacting systems, we will study the 1D single impurity Anderson model (SIAM) \cite{PhysRev.124.41}.
This is an ideal test system since the interaction is local and act only on the impurity site. The properties of this model are well 
understood, allowing us to physically interpret properties revealed by the transformations.

The Hamiltonian for the SIAM is
\begin{equation}\label{eq:H_A}
    H_\text{A} = H_{\mathrm{imp}} + H_{\mathrm{hyb}} + H_{\mathrm{bath}},
\end{equation}
with the impurity terms
\begin{equation}
\label{eq:H_imp}
    H_{\mathrm{imp}} = \epsilon_d (\hat{n}_{d\uparrow}+\hat{n}_{d\downarrow}) + U \hat{n}_{d\uparrow} \hat{n}_{d\downarrow},
\end{equation}
describing a degenerate impurity orbital occupied by $\hat{n}_{d\sigma} = d_\sigma^\dag d_\sigma$ electrons having spin $\sigma$, energy $\epsilon_d$, and an on-site Coulomb repulsion $U$.
The second term hybridizes the impurity with the first bath electron (denoted as site~$2$),
\begin{equation}
\label{eq:H_hyb}
    H_\text{hyb} = -V \sum_{\sigma} (d_\sigma^\dag c_{2\sigma} + \text{H.c.}),
\end{equation}
where $V$ is the hybridization strength.
The bath electrons are taken to be the 1D tight-binding model,
\begin{equation}
    H_\mathrm{bath} = -t \sum_{r=2,\sigma}^{N-1} (c_{r\sigma}^\dag c_{(r+1)\sigma} + \text{H.c.}),
\end{equation}
with hopping strength $t$ and total number of sites $N$, including the impurity.
For all cases, we keep the particle-hole symmetry for the impurity $U=-2\epsilon_d$. 

\begin{figure*}[t]
\begin{center}
\includegraphics[width = 0.8\textwidth]{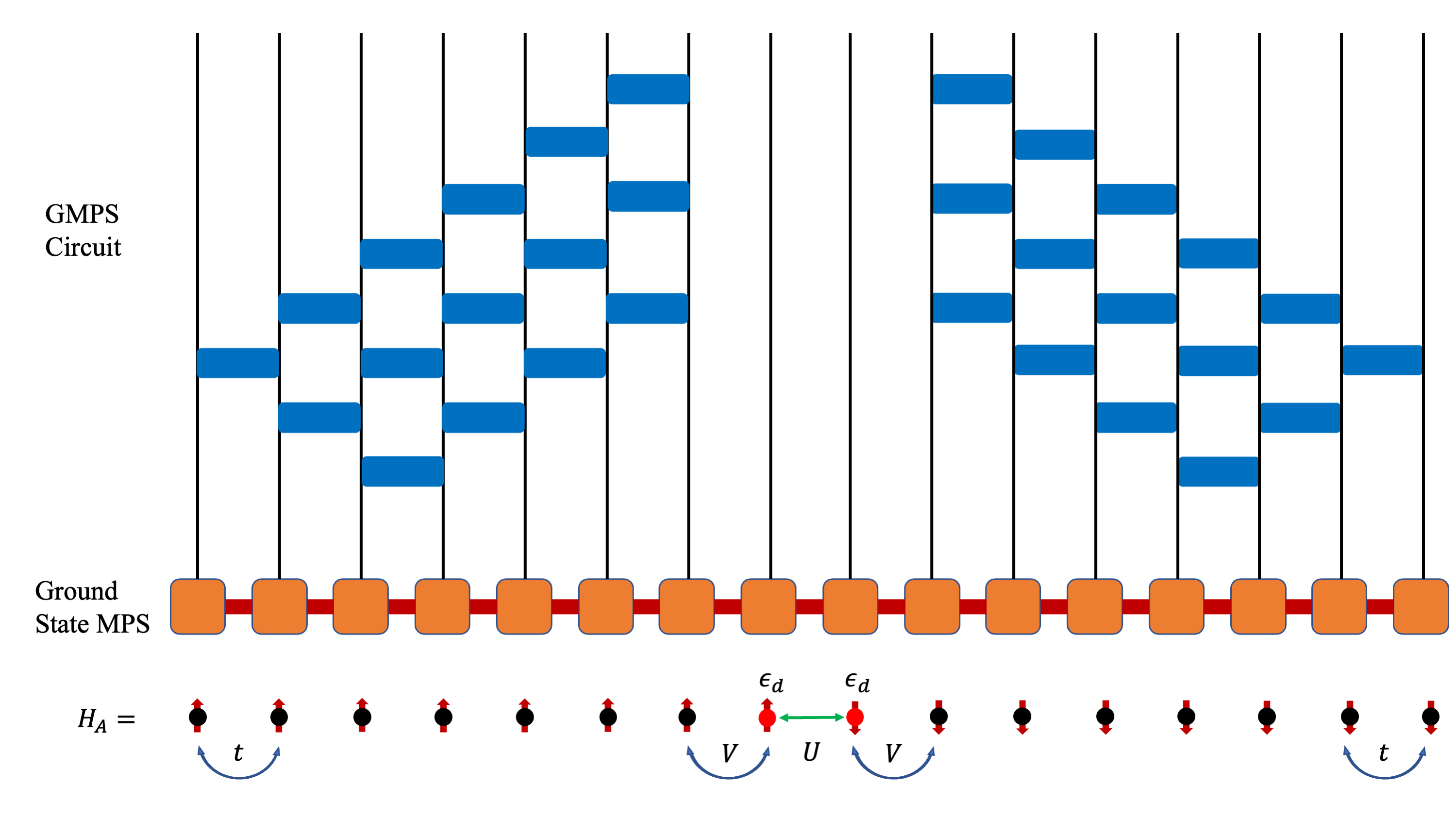}
\caption{\label{fig:model}   Demonstration of the split-site geometry of the impurity model and the corresponding GMPS circuit. In the bottom, the spilt-site geometry of the impurity model $H_\mathrm{A}$ with $N=8$ is shown with all up spin degrees freedom in the left-hand side while all down spins in the right with Eq. \ref{eq:splitop}. The corresponding GMPS circuit with block size $B=4$ can be obtained by symmetrizing the circuit from the up spin part (noninteracting limit). To maintain the locality of the interaction $U$, rotation gates at the impurity sites are avoided. Blocks of the circuit start from the two ends of the chain, each of which consist of $B-1$ unitary gates (blue rectangles).
}
\end{center}
\end{figure*}

Computationally, when using the spinful $S=1/2$ local Hilbert space $\ket{0}$,$\ket{\uparrow}$,$\ket{\downarrow}$,$\ket{\uparrow\downarrow}$ for each site in the original geometry of the model, the DMRG algorithm requires high bond dimensions $\chi$ (thousands for system sizes of hundreds) to obtain good precision. Without changing the physics, one can rewrite the model by separating the spin up and down operators,
\begin{equation}\label{eq:splitop}
    \begin{split}
        a_{-j} \equiv c_{j\uparrow}, \quad a_j \equiv c_{j\downarrow},
    \end{split}
\end{equation}
where $j=1,2,\cdots, N$ and $j=1$ denotes the impurity operator $a_{-1}\equiv d_\uparrow, a_1\equiv d_\downarrow$. Such a \emph{split-site representation} leads to more favorable entanglement and MPS bond dimensions.
This ``unfolded'' representation has been used often for studying impurity systems with MPS techniques \cite{Saberi:2008,Ganahl:2015,Rams:2020,Kohn:2021}. The geometry of this split-site representation and the corresponding GMPS circuit are shown in Fig. \ref{fig:model}. The detailed model in terms of operators $a_j$ can be found in Appendix \ref{app:splitsite}. Beyond the SIAM, we also consider two-impurity models, where we add an extra impurity at various locations with the same hybridization (see Sec. \ref{sec:twoimpurity}).

Throughout, we will be especially interested in studying 
the von Neumann entanglement entropy, defined as
\begin{equation}
\label{eq:entropy}
S = - \mathrm{Tr} \rho_A \ln\rho_A 
\end{equation}
where $\rho_A = \mathrm{Tr}_B\big[\ket{\Psi}\bra{\Psi}\big]$ is the reduced density matrix of system $A$ and $\ket{\Psi}$ is the quantum many-body state
of interest.

The MPS formalism allows one to obtain the entropy $S$ efficiently when $A$ is either a contiguous region of sites of the form $1,2,\ldots,L_A$
or when $A$ is a small region of neighboring sites. 
Computing the entanglement entropy of what would be a simple bipartitioning of the sites in real space (before splitting up and down spin degrees of freedom) 
involves a more complicated ABA partitioning in the split-site representation. 
We describe how to compute the entanglement entropy with MPS in the split-site representation in Appendix \ref{app:impurity-entropy}.

\section{Disentangling Impurity Models with Gaussian MPS Circuits}
\label{sec:disentangler}

\begin{figure}[t!]
\begin{center}
\setlabel{pos=nw,fontsize=\large,labelbox=false}
\xincludegraphics[scale=0.4,label=a]{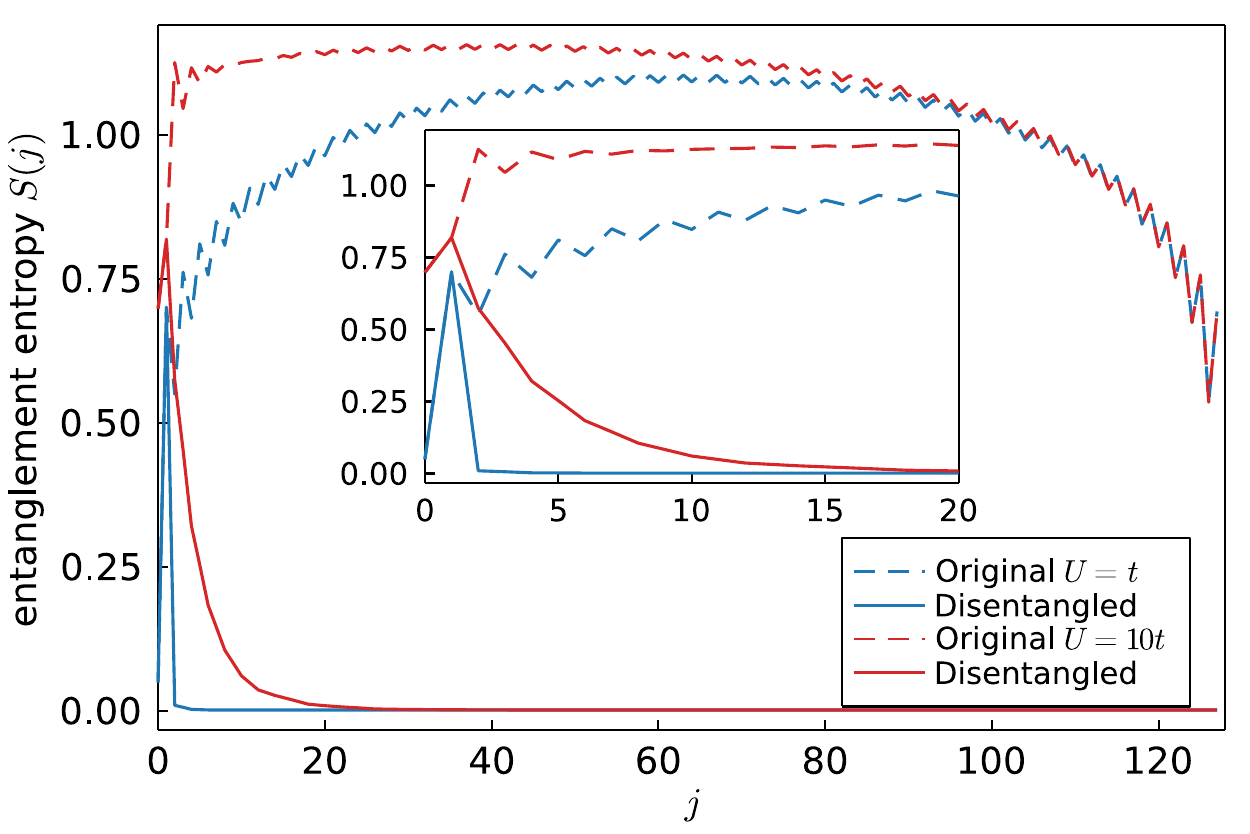}
\xincludegraphics[scale=0.4,label=b]{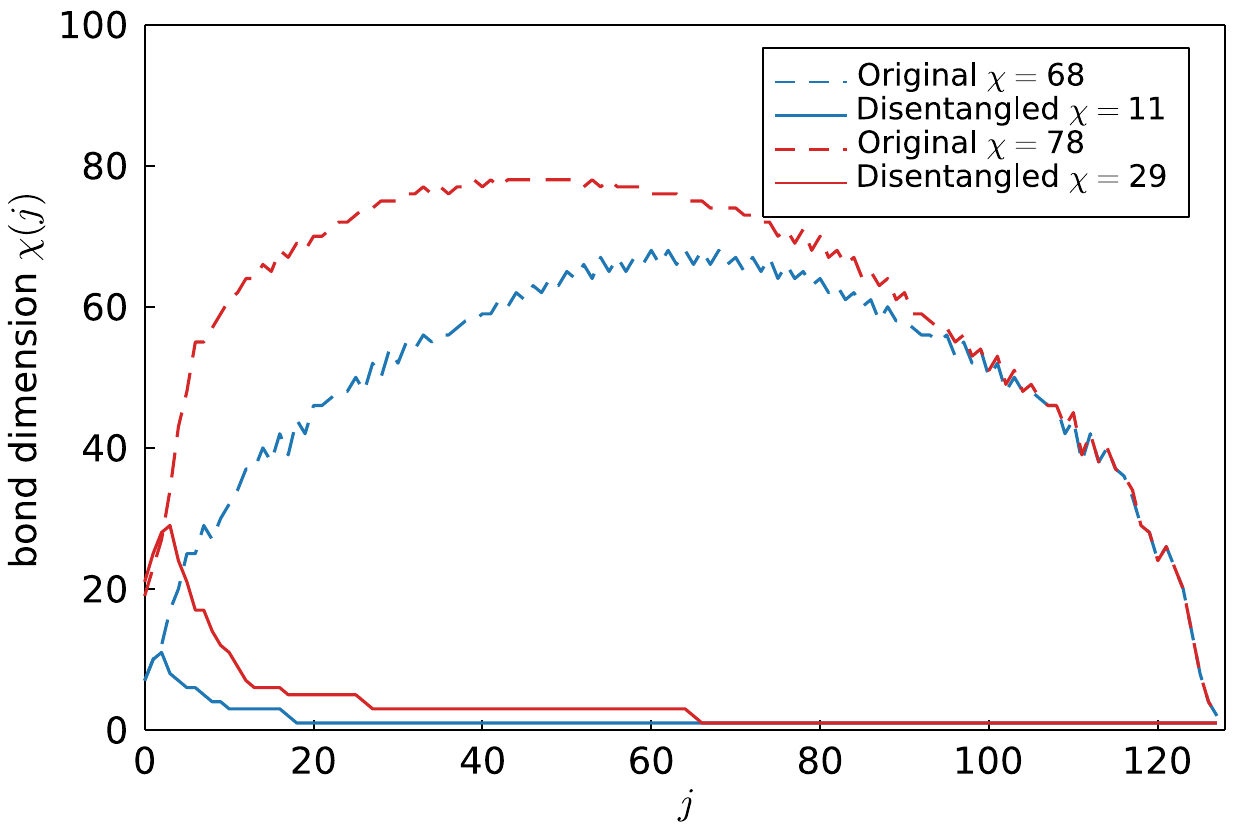}
\caption{\label{fig:dpsi} Entanglement entropy and MPS bond dimension before and after the fermionic Gaussian circuit transformation. The properties in the original basis (dashed curves) are compared with the same properties in the transformed basis (solid curves) when using the GMPS circuit for systems $N=128, V=t$. 
Systems with small interaction $U/t=1$  are shown in blue, and strong interaction $U/t=10$ in red.
Panel (a) shows the entanglement entropy $S$ as a function of bipartite (down spins) site distance $j$ away from the impurity site.
Inset: entanglement entropy close to the impurity. Panel (b) shows MPS bond dimensions $\chi(j)$ at each bond. The legend shows the maximum MPS bond dimension $\chi$ (when using a fixed truncation error cutoff in DMRG of $10^{-8}$). 
}
\end{center}
\end{figure}

We now explore the power of Gaussian MPS circuits for disentangling ground states. After disentangling an interacting ground-state MPS to demonstrate the idea, we use the circuit to change the basis of the Hamiltonian \emph{before} computing the ground state to accelerate tensor network calculations.

\subsection{Disentangling the ground state}

Using DMRG, we obtain the
ground state of the  single-impurity Anderson model (SIAM) as an MPS.
To disentangle this ground-state MPS, we compute a GMPS circuit from the $U=0$ limit of the SIAM, apply this circuit to the ground-state MPS, 
and study its entanglement  in the transformed basis.

Figure~\ref{fig:model} shows this process in more detail for the split-site geometry we used. The GMPS transformation starts from the ends of the finite leads
and proceeds toward the impurity. 
Any gates touching the impurity are omitted. Because of the left-right (up and down spin) symmetry of 
the model we use the same gates from both ends.

Figure~\ref{fig:dpsi}(a) shows the entanglement entropy before and after the transformation for both $U/t = 1$ and $U/t = 10$, showing only the down-spin (right-hand) side of the system.
The entanglement starts rather high on the original basis, with the characteristic ``arc'' shape of gapless systems. 
After the GMPS transformation, the entanglement is reduced nearly to zero over most of the system except
for a region near the impurity. Remarkably, the same circuit based on the $U=0$ limit works well in both cases.
The MPS bond dimension $\chi$ is also significantly reduced by the GMPS transformation [Fig.~\ref{fig:dpsi}(b)].

Certain physical properties become apparent after the transformation, a phenomenon we study in more detail in Sec. ~\ref{sec:screening}.
In the SIAM, the impurity is known to be screened by the bath electrons over a length scale $\xi_K$, the Kondo screening length, and should be primarily entangled with
sites within this distance. The Kondo screening length is known to become larger as $U/t$ increases.
Correspondingly, one can see in Fig.~\ref{fig:dpsi} that the region of finite entanglement
near the impurity is in fact larger in the $U/t=10$ system than in the $U/t=1$ system.

\subsection{Transforming the Hamiltonian}
\label{sec:MPO}

The speed of a DMRG calculation depends on the entanglement of the state being calculated.
But the approach of the previous section would not yield faster ground-state calculations since the state was computed before the transformation
was ever used. 
In this section, we exploit the dual nature of fermionic Gaussian circuits, as both many-body quantum circuits and single-particle
basis transformations \cite{Fishman:2015} to transform the Hamiltonian before starting any DMRG calculations. It is not obvious that this will work, as the more
complicated form of the transformed Hamiltonian could offset gains from the less-entangled state. But the tradeoff turns out to be very favorable.

To transform the Hamiltonian, we interpret the GMPS circuit as a ``matrix circuit'' of sparse Givens rotation matrices [Eq.~\ref{eq:Givens}] multiplied together to
form a unitary single-particle transformation $R$. Using the definition $f_r = \sum_{j} R_{r j} c_j$ of the fermion operators in the transformed basis, 
the quadratic part of the Hamiltonian transforms as
\begin{align}
\sum_{j j'} t_{j j'} c^\dagger_j c_{j'} 
& = \sum_{r r'} \Big(\sum_{j j'} R_{r j} t_{j j'} R^*_{r' j'} \Big) f^\dagger_r f_{r'} \nonumber \\
& = \sum_{r r'} \tilde{t}_{r r'} f^\dagger_r f_{r'}
\end{align}
As in the previous section, any matrices (or ``matrix gates'') in $R$ that act on the impurity are dropped from the circuit. 
This choice keeps the
impurity part of the Hamiltonian $H_\text{imp}$ precisely the same after the change of basis and
keeps the interaction $U$ term local.

\begin{figure}[t!]
\begin{center}
\includegraphics[width = 0.49\textwidth]{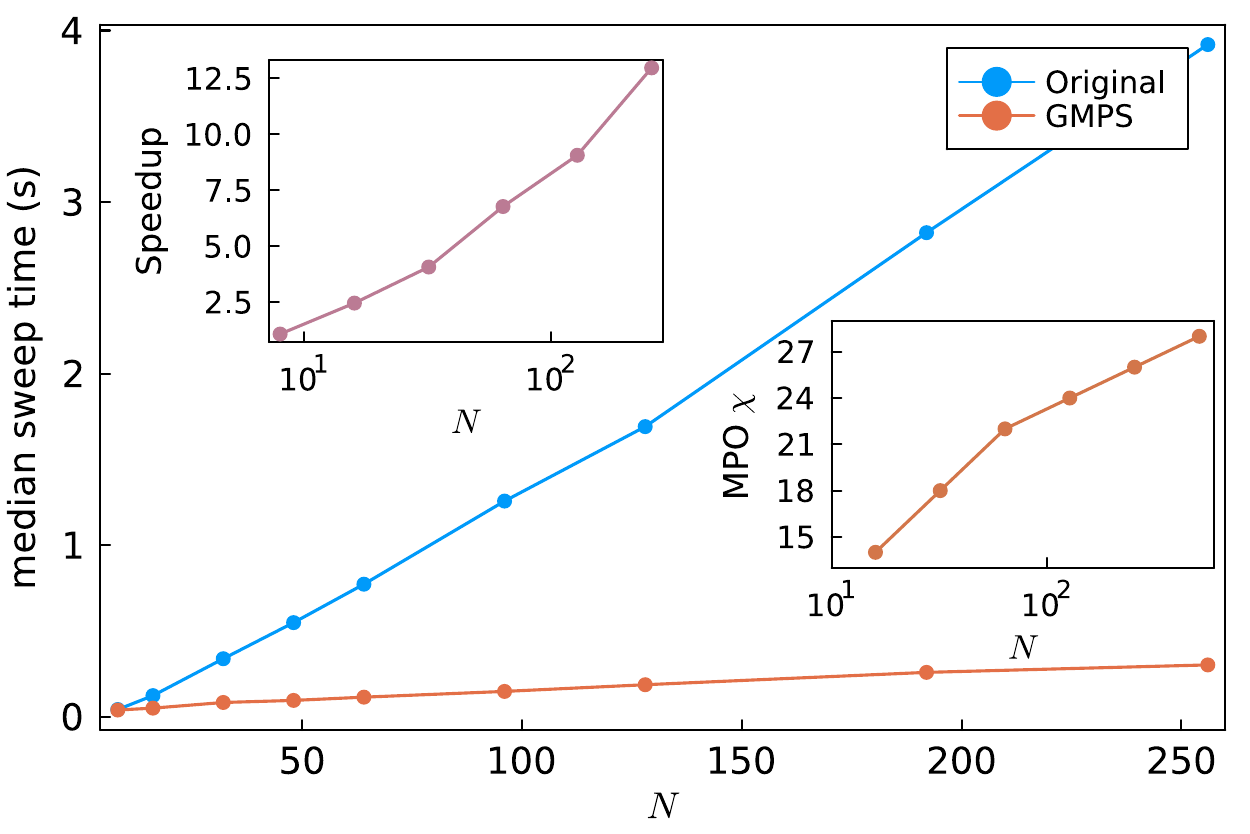}
\caption{\label{fig:without} Median DMRG sweep time in the original versus GMPS transformed basis as a function of system size $N$. Each DMRG calculation involved 20 sweeps and a truncation cutoff of $10^{-6}$; the system used was $U=t,V=t,\epsilon_d=-U/2.$  Top left inset: the speed-up, defined as the original sweep time divided by sweep time on the transformed basis, as a function of system size. Bottom right inset: the maximum bond dimensions $\chi$ of MPO as a function of system sizes $N$, which asymptotically scale as $\log(N)$. The sweep time was obtained from calculations on an M1 Macbook. To assess ground state energy convergence, we also performed more accurate DMRG calculations with a lower cutoff of $10^{-10}$. On the original basis, the corresponding energy change was $\Delta E=2.8\times 10^{-3}$ while it was $\Delta E = 1.5\times 10^{-5}$ on the GMPS basis.
}
\end{center}
\end{figure}

Running DMRG on the transformed Hamiltonian turns out to be much faster since the ground state has very low entanglement---see Fig.~\ref{fig:without} for timings.
The median sweep time is between 2.5 and 12.5 times \emph{faster} than in the original basis, with greater speedups for the larger the system size.

Because the Hamiltonian is less local in the transformed basis, the bond dimension of its 
matrix product operator (MPO) representation will necessarily become larger than in the original basis. 
To compute the Hamiltonian MPO in the transformed basis, 
we use a compression algorithm based on Ref.~\onlinecite{Chan:2016} that is publicly available through the 
ITensor software \cite{fishman2022itensor}.
The transformed MPO bond dimension grows logarithmically with system size (lower inset of Fig.~\ref{fig:without}),
reaching $\chi=26$ for the largest system size of $N=512$. This is in contrast to a bond dimension of $\chi=4$ in the original basis.
An MPO bond dimension of $\chi=26$ is not very large in practice, and DMRG calculations scale only quadratically with the MPO bond dimension,
versus as the cube of the MPS bond dimension, explaining why the ``disentangled'' calculations in the transformed basis can be so much faster
despite a more complicated Hamiltonian.

We conclude this section by mentioning an interesting negative result.
We attempted to reduce the MPO bond dimension further by discarding any quadratic
Hamiltonian terms with an absolute value below a small threshold $\tilde{t}_{ij} < \epsilon$.
But this led to \emph{larger} MPO bond dimensions compared to keeping all quadratic terms---see Appendix~\ref{app:MPOcutoff}
for details. In hindsight, this effect is probably due to the transformed $\tilde{t}_{ij}$ decaying smoothly, whereas truncating them would
introduce an artificially sharp step. Therefore, a recommended practice is to keep all Hamiltonian terms in the transformed basis when compressing the Hamiltonian into an MPO.

\section{Kondo screening physics in the transformed basis}
\label{sec:screening}

Gaussian circuit transformations remove entanglement by localizing ``inactive'' single-particle states. 
What physical properties become more transparent or easier to obtain in the transformed basis?
We will demonstrate that universal Kondo physics emerges from entanglement in the transformed basis and show that this basis is practically useful for computing other physical properties with high accuracy.

Kondo physics emerges from the SIAM in the low-energy limit, where impurity occupancy becomes pinned to $n_d = 1$ and a Schrieffer-Wolff
transformation maps the SIAM to an effective Kondo model in which the impurity couples to the bath only through a spin interaction (in the limit of particle hole symmetry). 
The strength of this spin interaction is called the Kondo (exchange) coupling and is given by 
\begin{align}
J_K = 8V^2 D/U,
\end{align}
with $D$ the half bandwidth for the particle-hole symmetric case. A renormalization group (RG) analysis of the Kondo model (in the metallic and thermodynamic limits) shows that the system always flows toward a strong-coupling fixed point where the impurity is screened by the bath electrons. The universal nature of the strong coupling Kondo screened fixed point implies that correlation functions, as well as the entanglement entropy, are scaling functions in terms of the Kondo screening length $\xi_K$ 
given by
\begin{align} \label{eq:xiK}
\xi_K \sim \frac{1}{\sqrt{U V^2}} \exp\Big(\frac{\pi U}{8 V^2}\Big) \ .
\end{align}
up to third order in $J_K$ within the RG
\cite{hewson1997kondo,PhysRevResearch.3.033188,coleman2015introduction}. 
In real space, this length scale can be associated with the size of the Kondo screening cloud.

\begin{figure}[t!]
\begin{center}
\setlabel{pos=nw,fontsize=\large,labelbox=false}
\xincludegraphics[scale=0.39,label=a]{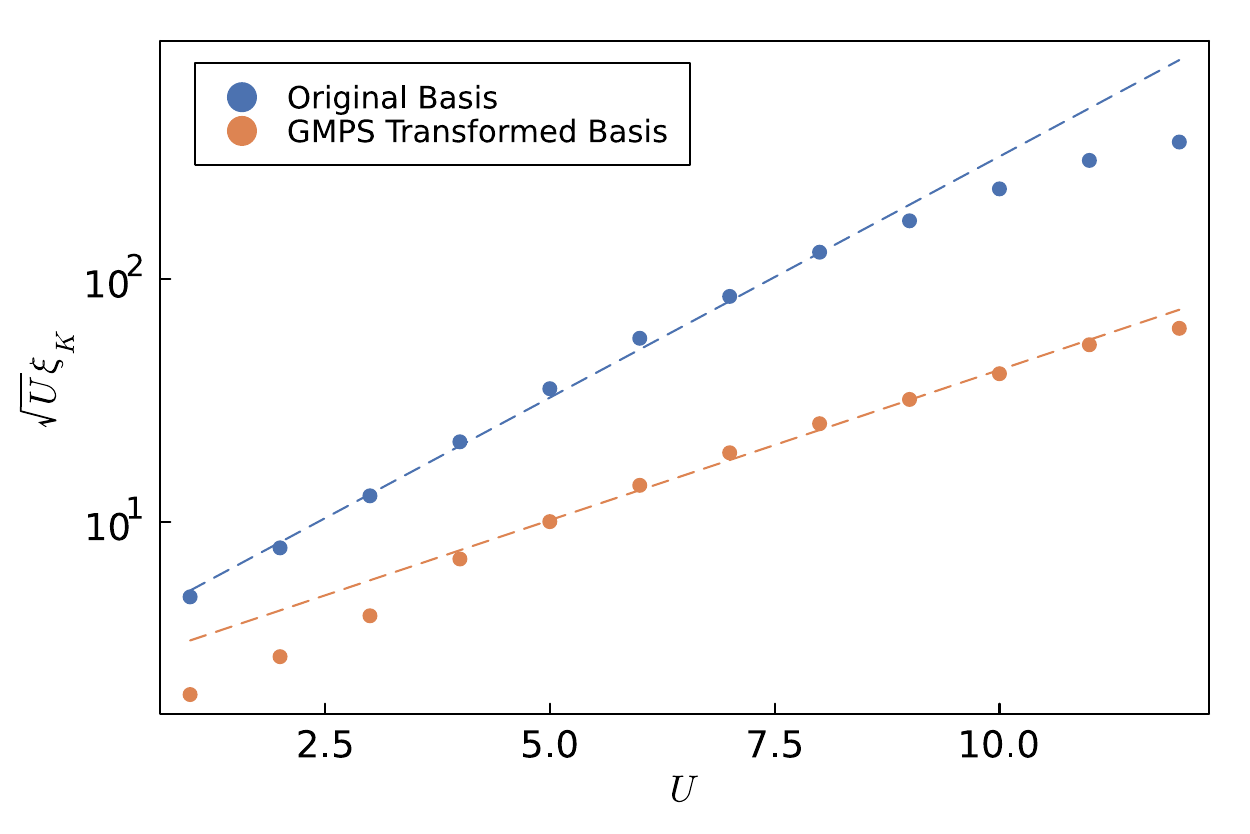}
\xincludegraphics[scale=0.39,label=b]{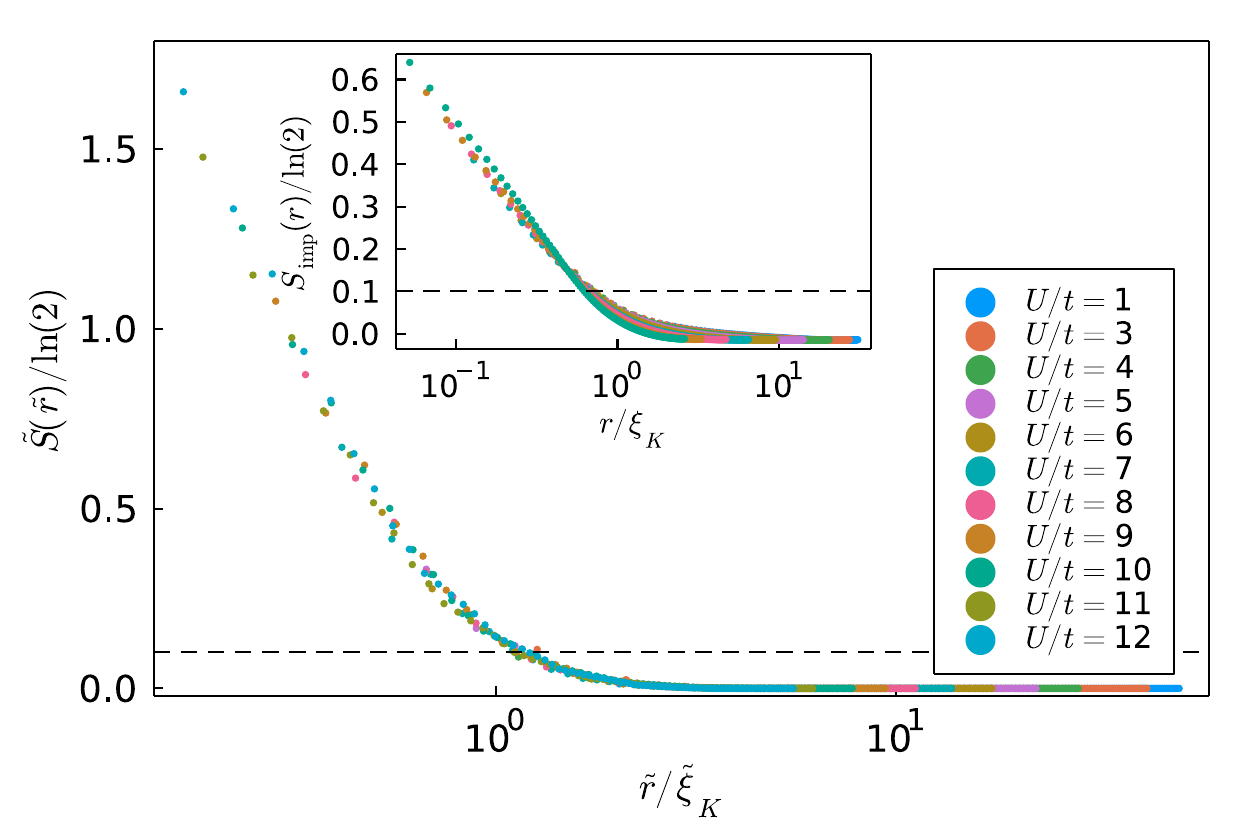}
\xincludegraphics[scale=0.39,label=c]{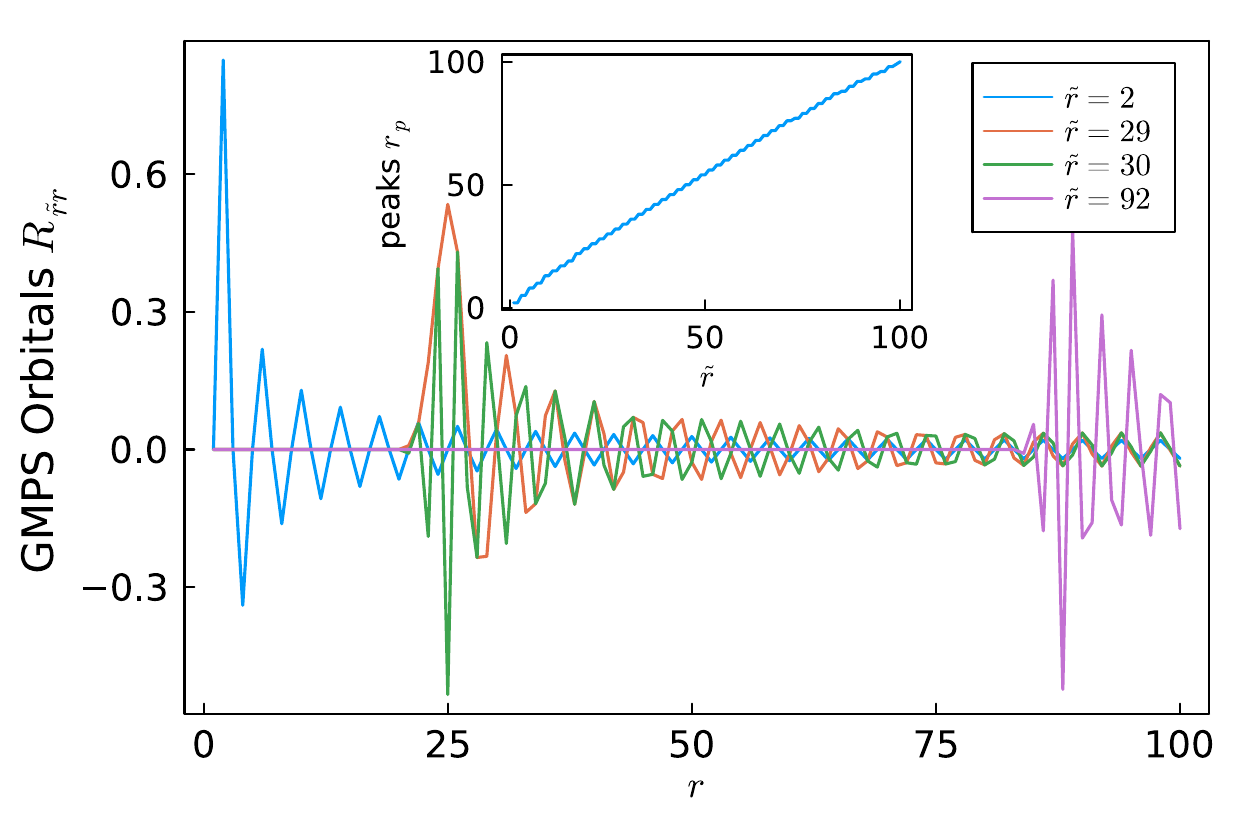}
\caption{\label{fig:S2} Impurity entanglement entropy and Kondo screening length from systems $N=400, V=t, \epsilon_d=-U/2$ with various $U \in [1, 12]$. (a) $\xi_K$, defined as the length where the entanglement entropy drops below the threshold $S_{\mathrm{imp}}(\xi_K) = 0.1$. (b) The collapse of entanglement entropy in the mapped GMPS basis, $\tilde{S}(\tilde{r})$, as a function of the scaled distance $\tilde{r}/\xi_K$ away from the impurity site, where $\xi_K$ are obtained from (a). Inset: corresponding results for impurity entanglement entropy $S_\mathrm{imp}(r)$ (with subtraction of bath electron contribution) in the original basis.
Note that in the original basis, we take middle locations across the even-odd result average. (c) Orbitals from the GMPS transformed basis with $N=100, V=t$. $r$ denotes the distance from the impurity $r=1$ in real space. Inset: location of the peaks $r_p$ in real space versus orbital index $\tilde{r}$ in the GMPS transformed basis.
}
\end{center}
\end{figure}

\subsection{Kondo Screening Length From Entanglement}

A  direct measure of the Kondo screening cloud is to study the behavior of the entanglement entropy at a distance $r$ away from the impurity \cite{PhysRevLett.121.147602,sorensen2007impurity,sorensen2007quantum,affleck2009entanglement,PhysRevResearch.3.033188,PhysRevB.91.245122}. 
It has been shown that the impurity entanglement entropy exhibits universal scaling behavior in terms of $r/\xi_K$. 
The impurity entanglement entropy in real space is defined as
\begin{equation}\label{eq:Simp}
\begin{split}
    S_{\mathrm{imp}}(r, U)&=S(r,U)-S^{(0)}(r), \\
    S(r, U) &= -\mathrm{Tr} [\rho_r \ln \rho_r],
\end{split}
\end{equation}
where $S^{(0)}(r)$ is entanglement entropy only from the bath ($U=0,V=0$), which can be obtained from the correlation matrix (Appendix A of Ref. \cite{Fishman:2015}), and $\rho_r$ is the reduced density matrix for the subsystem of the first $r$ sites around the impurity, computed from the interacting state at finite $U$. Computing the entanglement entropy in the split-site representation is explained in more detail in the Appendix \ref{app:impurity-entropy}.
The impurity entanglement entropy   oscillates between even and odd sites \cite{PhysRevResearch.3.033188,PhysRevB.53.9153, PhysRevLett.121.147602}, 
so in what follows we will always average over even and odd sites numerically. There exists a scaling function such that
\begin{equation}\label{eq:gx}
    S_{\mathrm{imp}}(r, U)= g\Big(\frac{r}{\xi_K(U)}\Big), 
\end{equation}
where $g(x)$ is a  universal scaling function. 

The purpose of bath subtraction in Eq.~(\ref{eq:Simp}) above is to reveal the location of the screening cloud, where deep inside the screening cloud $r\ll \xi_K(U)$, $S_{\mathrm{imp}}=\ln 2$, and outside of it $r\gg \xi_K(U)$,  $S_{\mathrm{imp}} \rightarrow 0$.
In the GMPS transformed basis, the decay of the entanglement away from the impurity is naturally achieved and a subtraction of the noninteracting bath entanglement is no longer needed (see Fig.~\ref{fig:dpsi}).
However, it is not obvious that the entanglement in the transformed basis will exhibit universal scaling behavior similar to
Eq.~\ref{eq:gx} since the distance $\tilde{r}$ from the impurity in the transformed basis is not the same as the real-space distance $r$.

Nevertheless, Fig.~\ref{fig:S2} shows we can obtain universal information about the Kondo screening physics directly from the entanglement in the transformed basis.
For the results in panel (a), $\xi_K$ is determined in the GMPS transformed basis by finding the value of $\tilde{r}$ such that $S(\tilde{r}=\xi_K) = 0.1$, where $0.1$ is an arbitrary small threshold. We also performed an identical procedure in the original real-space basis but using the quantity $S_\text{imp}$ Eq.~(\ref{eq:Simp}).
The values of $\xi_K$ determined this way scale with $U$ as $\sqrt{U}\xi_K\sim \exp(C U)$ over a wide range of $U$ with some constant $C$ (different for each basis) which fits qualitatively well with 
the expression in Eq.~\eqref{eq:xiK}.

In Fig. \ref{fig:S2}(b), we scale the $x$ axis by the Kondo length $\xi_K$ and plot the transformed basis
entanglement against this rescaled axis. There is a clear scaling collapse for a range of $U$, even better than for the real-space $S_\mathrm{imp}$ results
shown in the inset of Fig.~\ref{fig:S2}(b), and without the difficulty of subtraction and even-odd averaging necessary to analyze the entanglement collapse in 
real space \cite{PhysRevLett.121.147602}. To understand better what the $\tilde{r}$ distance means in the transformed basis, Fig.~\ref{fig:S2}(c) also shows the orbitals
$R_{\tilde{r}r}$ in real space $r$ corresponding to a specific site $\tilde{r}$ in the transformed basis.

\subsection{Universal Collapse of Spin Correlators}

Universal aspects of Kondo physics can also be realized through the correlations between the impurity spin and the noninteracting bath, given by
\begin{equation}\label{eq:correlator}
  C(r) =   \langle {\bf S}_d \cdot {\bf S}(r)\rangle=\langle \Psi_0 | {\bf S}_d \cdot {\bf S}(r)| \Psi_0 \rangle,
\end{equation}
where ${\bf S}_d = d^{\dag}_{\alpha}(\bm{\sigma}_{\alpha \beta}/2)d_{\beta}$ and $\mathbf{S}(r)=\frac{1}{2} \sum_{ss'} c^\dagger_{r\alpha} \boldsymbol{\sigma}_{\alpha \beta}c_{r\beta}$ is the impurity and conduction electron spin density, respectively, and $\bm{\sigma}$ denotes a vector of Pauli matrices. 
The spin correlations exhibit a universal scaling collapse of the form \cite{PhysRevB.75.041307,PhysRevB.80.205114,PhysRevB.57.432}
\begin{equation} \label{eq:fx}
    C(r, U)= \frac{1}{\xi_K}f\left(\frac{r}{\xi_K}\right),
\end{equation}
where $f(x)$ is a universal scaling function. 

\begin{figure}[t!]
\begin{center}
\setlabel{pos=nw,fontsize=\large,labelbox=false}
\xincludegraphics[scale=0.4,label=a]{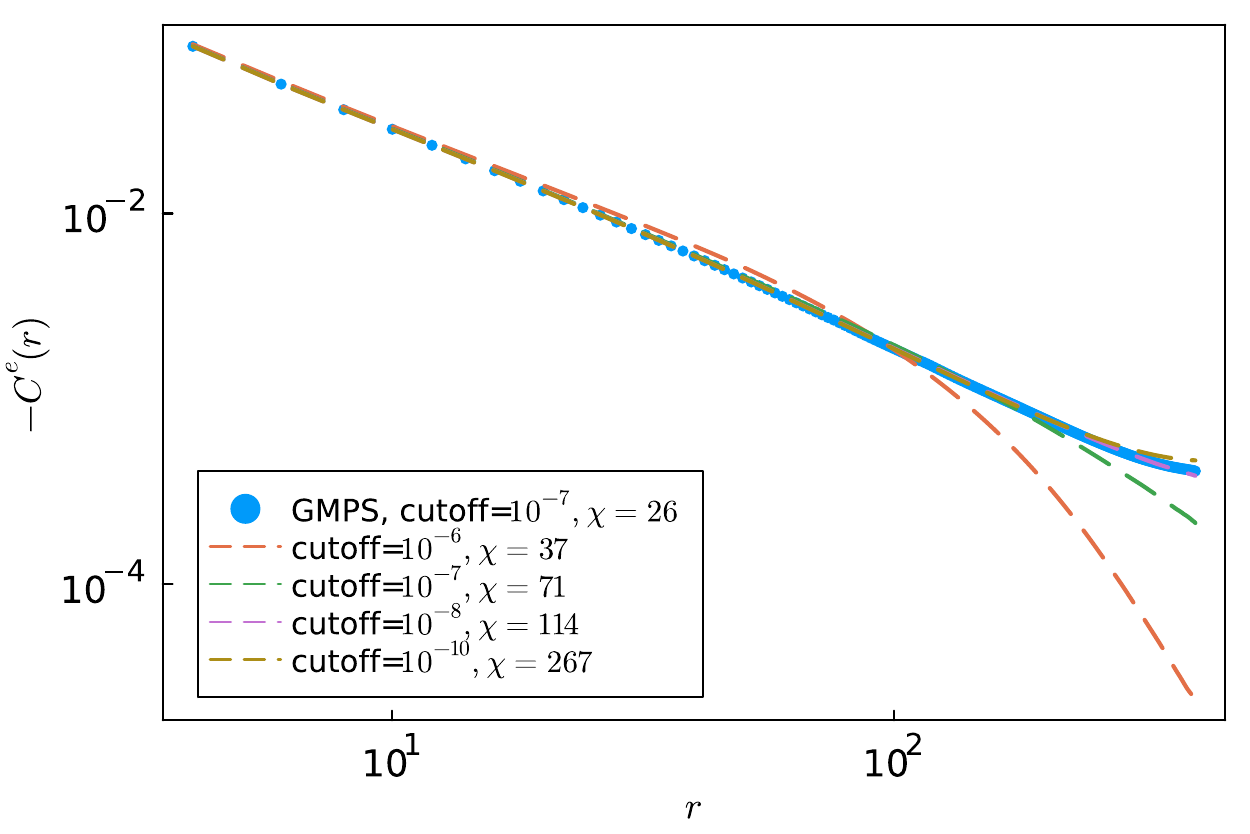}
\xincludegraphics[scale=0.4,label=b]{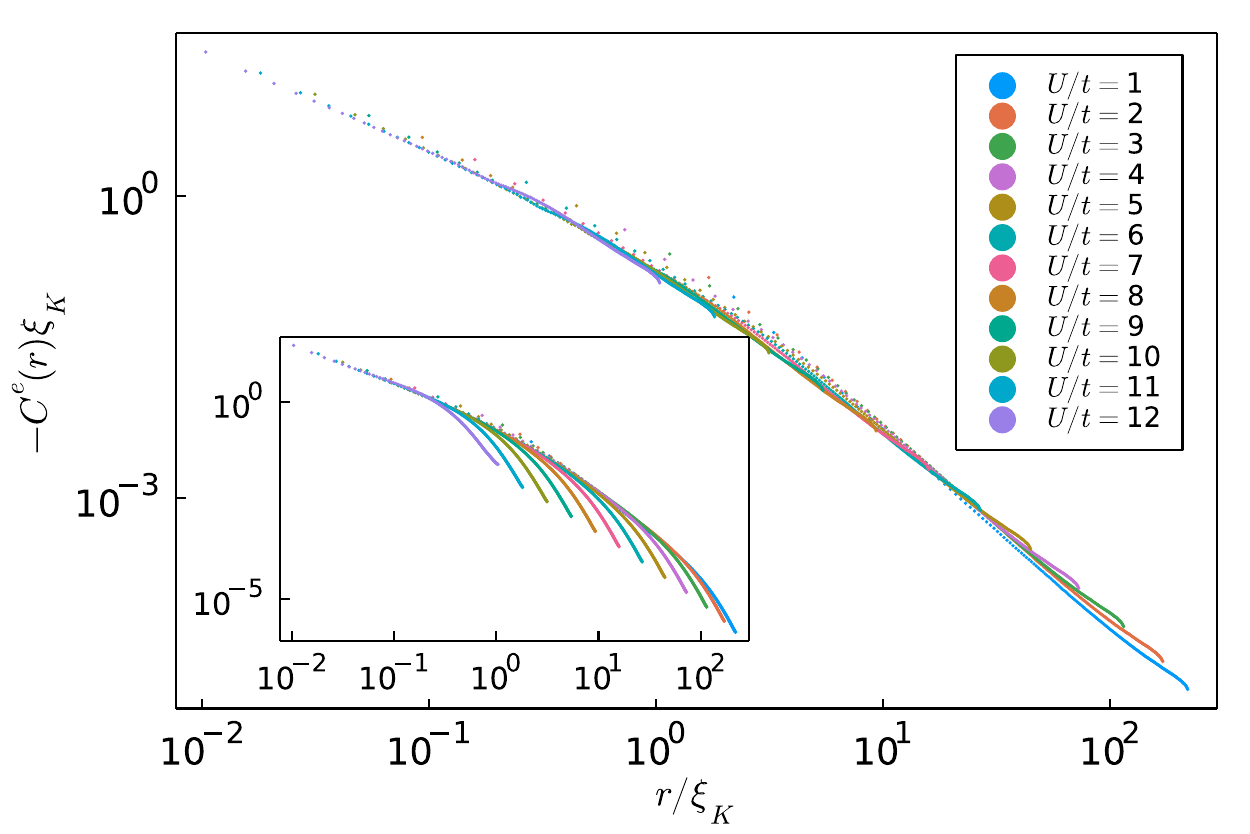}
\caption{\label{fig:Cr} Even site spin-spin correlator $C^e(r)$ (Eq. \ref{eq:correlator}) and the scaling collapse for $N=400, V=t, \epsilon_d=-U/2$ with various $U \in [1, 12]$. Empirically, the screening length is found to be $\xi_K =e^{0.6U}/\sqrt{U}$. 
(a) The correlation function $C(r)$ with $U=10t$ from the DMRG in the GMPS transformed basis (sold circles) and in the original basis with various DMRG calculation cutoffs (dashed curves). 
(b) The correlation function $C(r)$ first obtained from the DMRG algorithm in the GMPS transformed basis and then transformed back to the original basis with the DMRG calculation cutoff$=10^{-6}$. Inset: results in the original basis with the same axes, parameters and DMRG calculation cutoff. 
}
\end{center}
\end{figure}

In Fig.~\ref{fig:Cr}(a) we show the spin correlation function $C(r,U=10t)$ computed in real space for various
DMRG truncation error cut-offs (smaller cut-off means higher accuracy) or in the transformed basis using a moderate
cut-off. We show results for even sites only to avoid oscillations. To obtain the transformed-basis results, we first converge the DMRG calculation, apply the GMPS circuit to map back to the original basis, then compute measurements. From the figure, we can see that the transformed-basis approach
produces results matching the highest-accuracy calculation in the original basis with a much more moderate
computational effort (smaller bond dimension). 

The advantage of performing computations using the transformed basis becomes even more noticeable when
we consider a range of interactions $U$ in Fig.~\ref{fig:Cr}(b) and plot the spin correlations with the axes
rescaled by $\xi_K$ to obtain a universal scaling collapse, using a value of $\xi_K$  determined analytically from Eq.~\ref{eq:xiK}. The main figure shows the excellent collapse obtained for
$1 \leq U/t \leq 12$ when using a DMRG cutoff of $10^{-6}$ and the transformed-basis approach. The inset
shows a similar calculation but performed entirely on the original basis, exhibiting noticeable deviations from the universal
form at higher values of $r$.

\section{RKKY effect in the two impurity model}\label{sec:twoimpurity}

We have shown that Gaussian fermionic circuits give a better computational basis for single-impurity models with local orbitals, where the interaction is local at the impurity site.  In this section, we discuss the behavior of the circuits in systems with more impurities at various locations in a one-dimensional chain. In the single-impurity case, we could conveniently work with spinless fermion operators by spatially separating up and down electrons and putting impurity in the middle of the chain, with the benefit that the MPS bond dimensions are approximately square root of the original spinful system. However, when introducing a second impurity, we cannot use the same trick and so we need to directly translate Gaussian gates into spinful electron space. The two-body rotational gates becomes $16$ by $16$ matrices instead of $4$ by $4$ matrices, where the new annihilation operator at site $j$ is $d_{j\sigma}= \cos\theta c_{j\sigma}-\sin\theta c_{j+1\sigma}$. 

\begin{figure*}[t]
\centering
\includegraphics[width=0.95\textwidth]{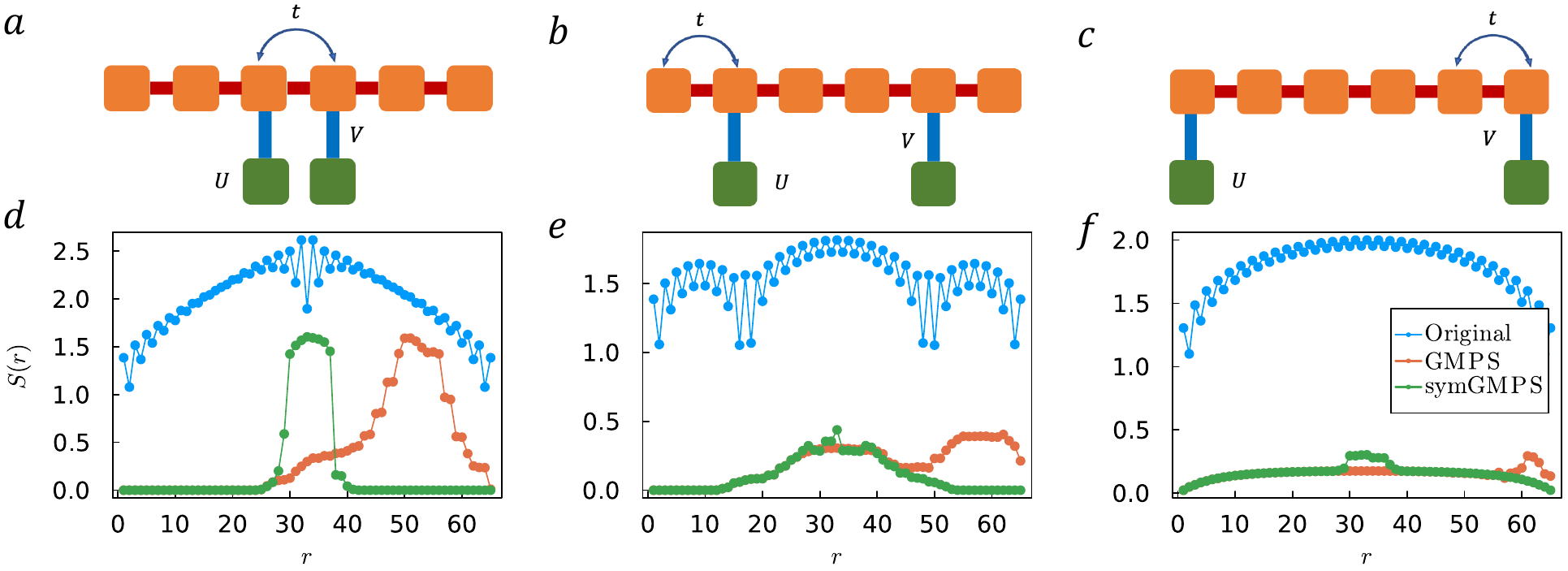}
\caption{\label{fig:twoImp} The two-impurity models with various impurity geometries. (a--c) Different symmetric geometries of two impurities in a one-dimensional chain. Orange squares represent conduction electron sites and green squares are impurity sites. The subfigures show (a) Two impurities in the center. (b) Two impurities at one and three quarters of the chain. (c) Two impurities on two ends. (d--f) Entanglement entropy of the corresponding ground states for each geometry in panels (a--c) with $U=t, V=t, L=66$. 
}
\end{figure*}

We consider geometries where the two impurities are placed symmetrically [Figs. \ref{fig:twoImp}(a)-\ref{fig:twoImp}(c)]. To make the construction of the GMPS circuits symmetric around the center of the chain, the circuits are built iteratively from the left and the right ends of the system, which we call the symGMPS approach as shown in Fig. \ref{fig:twoImp}(d)-\ref{fig:twoImp}(f). From Fig.~\ref{fig:twoImp}, the entanglement entropy correlates the two impurities based on their separation on the basis of symGMPS. When the two impurities sit in the center, there is a string of finite entanglement around the center. When the impurities are located at the ends, the string of entanglement extends to the whole chain, which we interpret as preserving correlations of the two impurities across the system. When the impurities sit at one and three quarters of the chain, the string expands approximately the real-space separation of the impurities. The symGMPS basis provides a novel way to show the impurity-impurity correlation in terms of residual entanglement entropy (entropy not removable by the Gaussian transformations), which reveals the emergence of the RKKY interaction \cite{PhysRev.96.99,10.1143/PTP.16.45,PhysRev.106.893}. The computational advantage of Gaussian transformations for these two-impurity models is less obvious, and it depends on both  the geometry and interaction strength. In Appendix \ref{app:twoImpSpeedup}, we demonstrate that in the side-geometry [Fig. \ref{fig:twoImp}(c)] case, the disentangled bases do provide computational benefits.

\section{Time evolved states transformed by fermionic Gaussian circuits}
\label{sec:timeEvolution}

An interesting question is whether we can use GMPS circuits  computed from noninteracting ground states to reduce the entanglement of systems undergoing dynamics.

To understand the dynamics of the impurity, we consider time evolution of state with a local quench at the impurity
site at time $t=0$
\begin{equation}\label{eq:quench}
    \ket{\Phi(t)}\equiv U(t)\ket{\Phi(0)}, \quad \ket{\Phi(0)}\equiv d_\downarrow\ket{\Psi_0}
\end{equation}
where $\ket{\Psi_0}$ is the many-body ground state and $U(t)=e^{-iHt}$ is the time-evolution operator. Since we are considering the system at half-filling ($U=-\epsilon / 2$), the dynamics for both spins are identical. 
The dynamics of such a locally quenched state can be used to compute the impurity Green's function. In particular, we compute the lesser Green's function
\begin{equation}
    iG^{<}(t)= \langle d^\dagger_{\downarrow}(t)d_\downarrow(0)\rangle =\bra{\Psi_0}U^\dagger(t) d^\dagger_\downarrow U(t)d_\downarrow\ket{\Psi_0}.
\end{equation}
Note that we can calculate the Green's function at twice the evolved time $t$ (time-doubling trick, Appendix \ref{app:timedoubling}),
\begin{equation}
    iG^{<}(2t)=\exp(2iE_0t) \langle{\Phi(-t)}|\Phi(t)\rangle,
\end{equation}
with the ground state energy $E_0$, although we only make use of this expression in the following to compute reference results. We compute the dynamics in both the interacting ($U/V=4$) and noninteracting case using the time-dependent
variational principle applied to MPS (MPS-TDVP)\cite{PhysRevLett.107.070601,PhysRevB.94.165116} with a timestep of $\delta t=0.1$. In particular, we employ the single-site formulation of MPS-TDVP with a subspace expansion that guarantees convergence to the result of the two-site formulation \cite{KlossInPrep}
, and we choose parameters that are equivalent to a discarded weight $\epsilon=10^{-14}$.

\begin{figure}[t!]
\begin{center}
\includegraphics[width=\columnwidth]{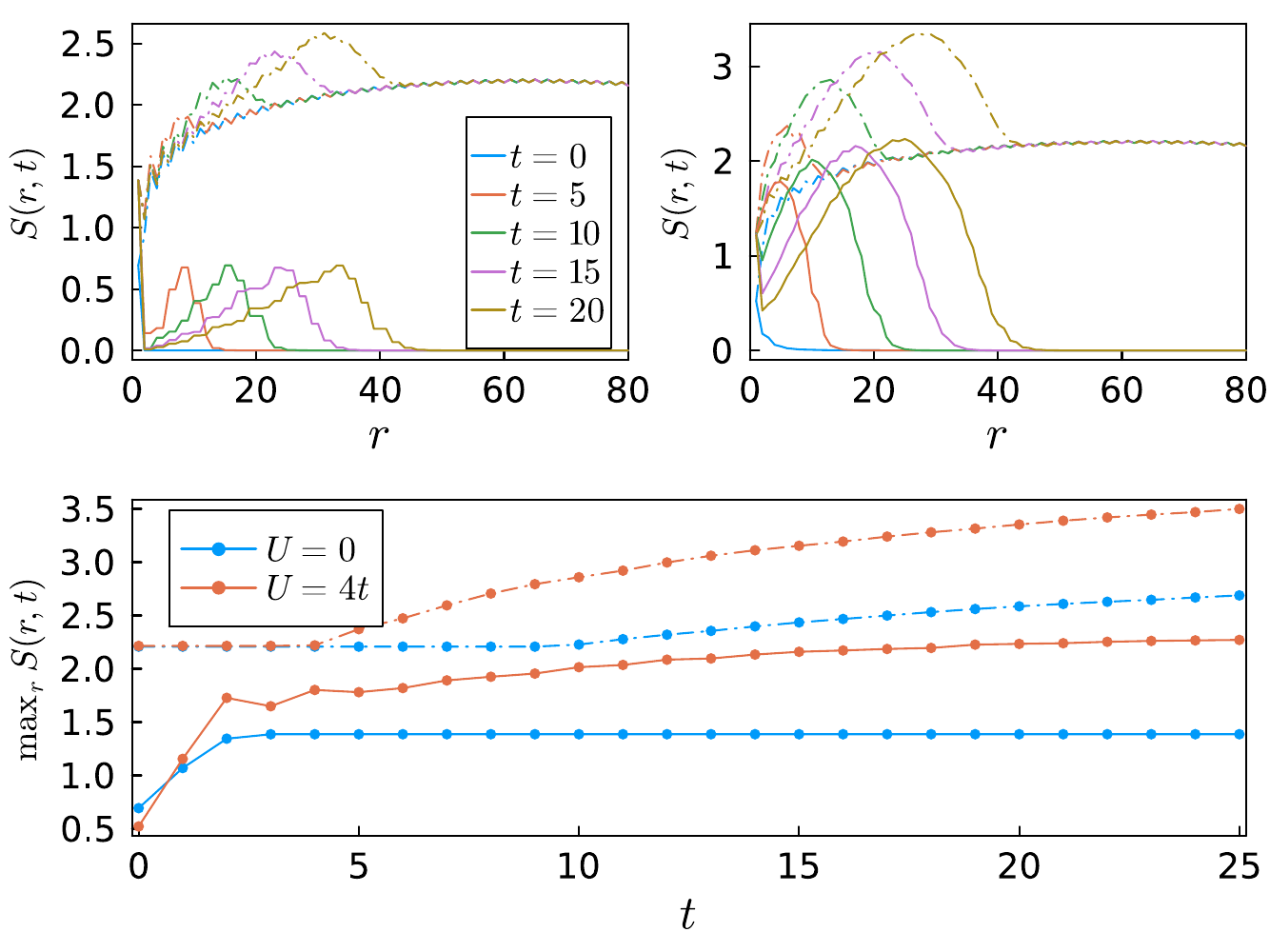}

\caption{\label{fig:tevo}  The entanglement entropy of the quenched system for $N=128, V=1.0t$. Solid curves are the entanglement after applying the GMPS circuit (disentangled), while dash-dotted curves are the original basis results. Upper panels: The entanglement entropy as a function of the bipartite (real-space) sites for several times for $U/V=0.0$ (left panel) and $U/V=4.0$ (right panel). (b) The maximum entanglement entropy (across sites) as a function of time. 
}
\end{center}
\end{figure}

\begin{figure}[t!]
\begin{center}
\setlabel{pos=nw,fontsize=\large,labelbox=false}
\includegraphics[width=\columnwidth]{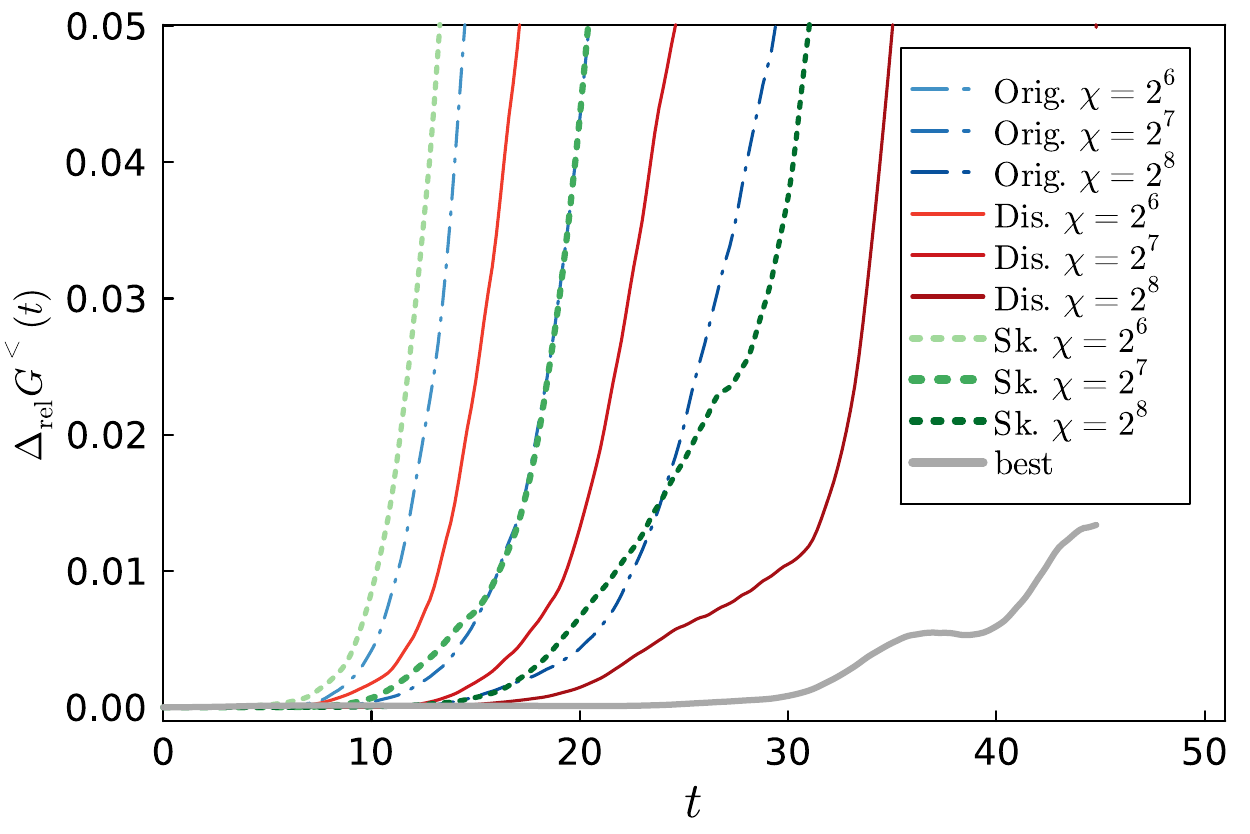}
\caption{\label{fig:Glesser_dev}  Relative deviation of the (negative) imaginary part of the lesser Green's function for various bond dimensions (color darkness) and bases (line styles). The reference calculation uses $\chi=2^9$, and the line labeled as ``best'' denotes the difference between the reference calculation with and without using the time-doubling trick in the original basis. The bases are: original (Orig.), disentangled via GMPS circuit (Dis.) and the bases of Ref. \cite{Kohn:2021} (Sk.)
}
\end{center}
\end{figure}

\begin{figure}[t!]
\begin{center}
\setlabel{pos=nw,fontsize=\large,labelbox=false}
\includegraphics[width=\columnwidth]{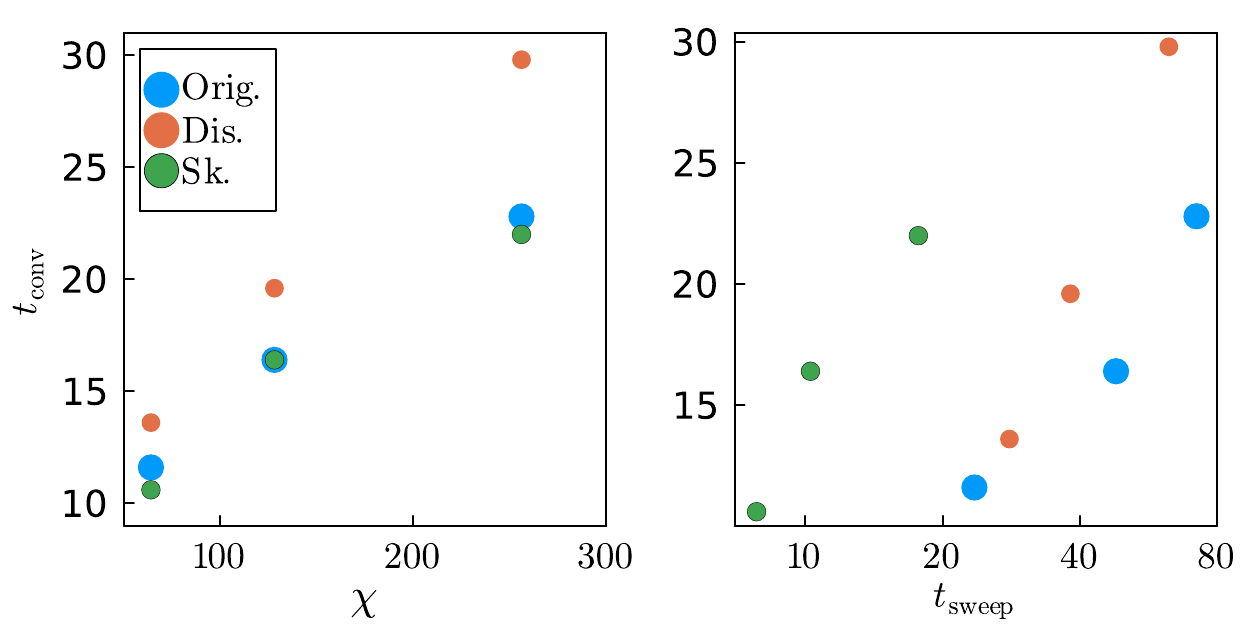}
\caption{\label{fig:convergence_times}  Maximum time up to which results remain accurate. The panels show the time up to which the lesser Green's function is converged within one percent versus bond dimension (left panel) and averaged wall time per time step within the converged time (right panel), excluding first step, for various bases (colors) from the same data as in Fig. \ref{fig:Glesser_dev}.
}
\end{center}
\end{figure}

The physical entanglement entropy for the quenched state is $S(r, t) = -\mathrm{Tr}[\rho_r(t) \ln \rho_r(t)]$, 
where $\rho_r(t) = \mathrm{Tr}_{\bar{r}} \rho, \rho = \ket{\Phi(t)}\bra{\Phi(t)}$ and $\mathrm{Tr}_{\bar{r}}$ denotes trace over complement physical (spinful electron) sites that are a distance of $r+1$ to $N$ away from the impurity site (see Appendix \ref{app:impurity-entropy} for details of calculation in \emph{split-site} representation) \footnote{The spatial index refers $r$ to a spatial orbital in the single-particle basis, i.e.~both spin-orbitals are labeled with the same spatial index. Due to the use of the \emph{split-site} representation which spatially separates the spin degrees of freedom in either half of the lattice in the MPS calculations, this entanglement entropy is not directly related to the singular values of the MPS in the canonical gauge}.

Figure ~\ref{fig:tevo} shows the behavior of the entanglement entropy for the time-evolved quenched state $\ket{\Phi(t)}$. The upper panel shows the spatial entanglement structure for a various times for both the interacting and noninteracting cases. Note that at time $t=0$,
applying the GMPS circuit removes nearly all of the entanglement from the quenched state $\ket{\Phi(0)}$ in both cases.
As the state evolves, the transformed-basis entanglement grows outward from the impurity and spreads across further sites. 
Interestingly, the region where the transformed-basis entanglement is nonzero corresponds to where the original-basis entanglement deviates from its time $t=0$ value. This is consistent with the disentangling transformation being local (finite-range). In both the interacting and noninteracting case, the transformed basis significantly lowers the maximal entanglement entropy at a given time. In the absence of interactions, the maximum entanglement entropy in the transformed frame is constant in time. Otherwise, the entanglement entropy shows a sublinear growth with time, in agreement with the logarithmic growth of entanglement entropy following a local quench predicted by conformal field theory \cite{Calabrese_2007}. 

The numerical value of the entanglement entropy is only a lower-bound for the MPS bond dimension, which is rarely saturated under unitary time-evolution from a weakly entangled state. It is thus not necessarily a good indicator of the required computational resources to represent a state faithfully as an MPS. Ultimately, we are interested in the accessible time-scale given a computational budget, i.e., a maximum fixed bond dimension or maximum total calculation time, under the condition that the result be converged within some tolerance.

Turning now to the accuracy of the impurity Green's function, we report 
the relative deviation from a reference calculation for the interacting case in Fig.~\ref{fig:Glesser_dev}. 
Since the Green's function $iG^<(t)$ oscillating around zero with a decaying envelope, we compute the relative deviation with respect to the smooth envelope obtained by Gaussian broadening applied to the absolute value of the Green's function. We also compare with the single-particle basis introduced by Santoro and Kohn \cite{Kohn:2021}. The latter also reduces the entanglement far away from the impurity for the interacting ground state, but is less effective in suppressing the entanglement in proximity of the impurity. For the quench dynamics, we find that the transformed basis obtained from the GMPS circuit gives superior results when working with a fixed maximum bond dimension compared to the other basis choices. Note that all calculations are performed with a very small discarded weight $\epsilon$ ($\epsilon=10^{-12}$ for the original basis, and $\epsilon=10^{-14}$ for the other bases) which leads to relatively quick saturation of the bond dimension to its maximum allowed value. We found that converging the calculation with respect to the discarded weight instead gives worse results.

To analyze why the Gaussian disentangled basis is able to deliver better accuracy for fixed-bond-dimension states, 
we extract $t_{\mathrm{conv}}$, the time up to which the result is converged with respect to reference within one percent. In Fig.~\ref{fig:convergence_times}, we show how $t_{\mathrm{conv}}$ behaves as a function of maximum bond dimension as well as the average ``wall time'' per time step, calculated in the range $t\in(0,t_{\mathrm{conv}}]$. For a given bond dimension, the disentangling of the GMPS circuit gives the largest accessible timescale, followed by the original basis (real space with split-site representation) and the basis of Ref. \cite{Kohn:2021}. The picture is more nuanced when we consider the reachable timescales for a given computational effort. At small to intermediate $t_{\mathrm{conv}}$, the basis of Ref.~\onlinecite{Kohn:2021} allows for a much faster calculation than the other two choices. However, the disentangling transformation obtained from the GMPS circuit provides access to the largest $t_{\mathrm{conv}}$ with a smaller effort than the original basis. Importantly, the wall time of the GMPS circuit calculation also appears to grow more slowly than the one using the basis of Ref. \cite{Kohn:2021}, which leads us to expect an advantage in the long-time limit.

Our results demonstrate that fermionic Gaussian circuits allow us to gain a more compact representation of quench dynamics even in the presence of moderately strong interactions. In terms of computational efficiency, both our basis and the one of Ref. \cite{Kohn:2021} are advantageous over the original basis. Since the Hamiltonian has a much smaller bond dimension in the basis of Ref. \cite{Kohn:2021} than in that of the GMPS circuit, it outperforms the other bases at shorter times. Arguably more important is the performance at late times, where the more effective disentangling of the GMPS circuit appears to outweigh the overhead of its larger Hamiltonian bond dimension. We also note that future algorithmic optimizations may allow us to decrease the prefactor associated with larger Hamiltonian bond dimension calculations.

Another advantage of the GMPS circuit over basis constructions that rely on physical intuition regarding the original problem is the potential for generalization. For example, one could imagine to construct GMPS circuit transformations based on the time-evolved state's single-particle density matrix (which ceases to be Gaussian) to reduce the entanglement as we time evolve the state. Whether this is feasible and leads to a computational advantage should be investigated in future work.

\section{Entanglement Renormalization of the Bath Electrons}
\label{sec:GMERA}

\begin{figure}[t!]
\begin{center}
\includegraphics[width = 0.49\textwidth]{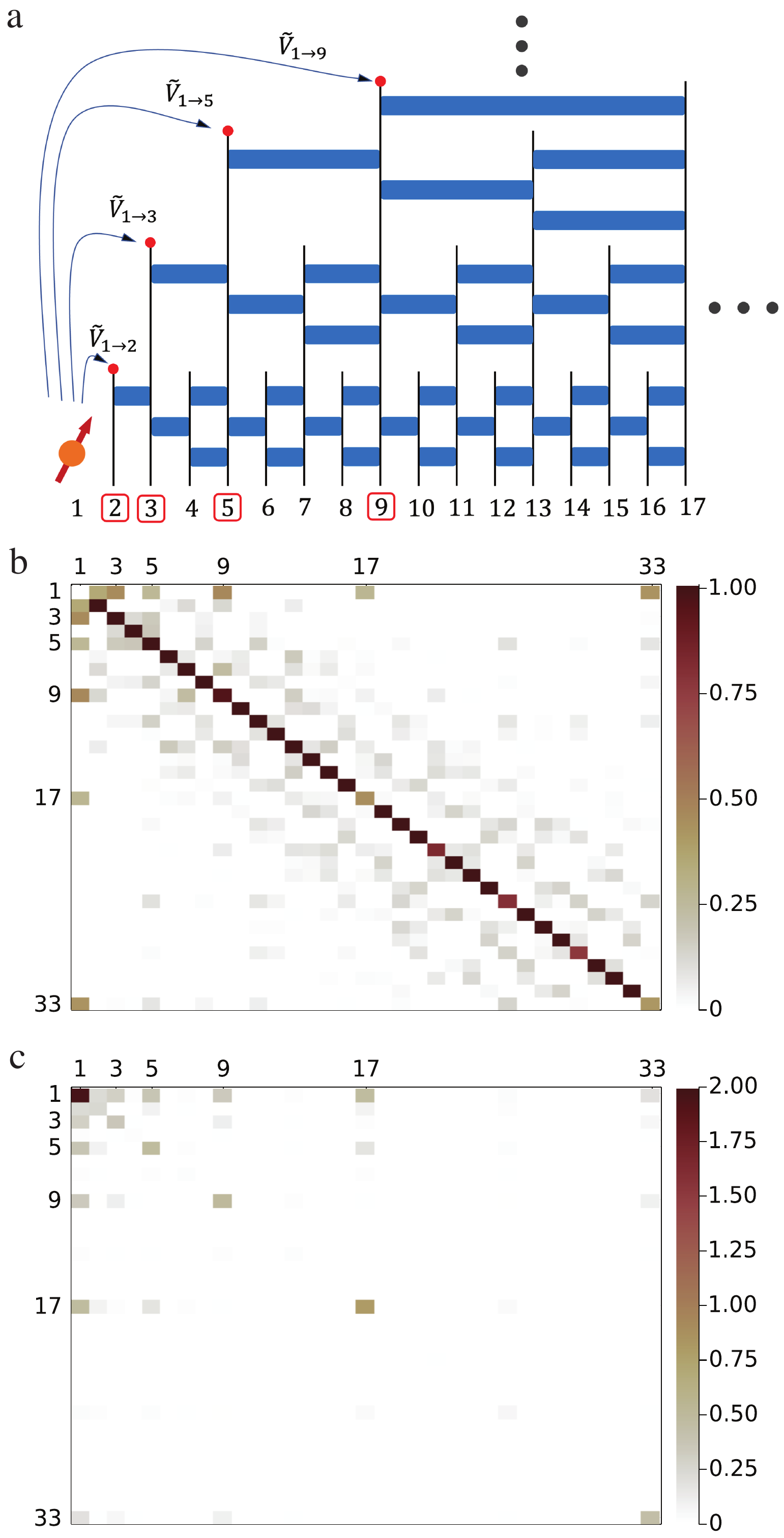} 
\caption{\label{fig:GMERAdemo}  Demonstration of the GMERA circuit. (a) The GMERA circuit obtained by diagonalizing odd sites in each iteration with system size $N=16$ and block size $B=4$. The impurity acts as a local spin in the left. The GMERA circuit generates effective hybridizations from the impurity to exponential sites, $r=2^k+1, k=0,1,\cdots$, marked by the red dots and red squares. (b) The transformed hopping Hamiltonian ($U=0$) under the GMERA circuit from the bath electrons with $N=80, V=t$, where the effective hybridizations (color bar) are prominent. (c) The part of mutual information matrix $I(r_1,r_2)$ of the ground state with $U=t$ for $N=80, V=t$. The peaks (color bar) in the mutual information perfectly fit with exponential sites appeared in panels (a) and (b).
}
\end{center}
\end{figure}

In the sections above, we considered fermionic Gaussian circuits with the structure of the GMPS, which worked very well in disentangling the interacting problem with local orbitals. However, the large degeneracy of the correlation matrix eigenvalues gives us freedom to choose other diagonalization procedures. In this section, we introduce other types of fermionic Gaussian circuits with structures inspired by the multiscale entanglement renormalization ansatz (MERA) and discuss how the transformed Hamiltonians have interesting emergent properties in the context of finding ground-state energies and performing time evolution.

\subsection{GMERA and Emergent Models}
\label{subsec:GMERAGS}

By introducing nonlocal unitary gates with a hierarchical structure, a class of fermionic Gaussian circuits implementing the MERA tensor network structure were proposed in Ref.~\onlinecite{PhysRevB.81.235102} and a simplified construction was introduced in Ref.~\onlinecite{Fishman:2015}. This fermionic Gaussian circuit is called the Gaussian MERA (GMERA) circuit, and its construction can be conveniently realized as a simple modification of the GMPS circuit construction where the gates are reorganized into a hierarchical structure. The GMERA circuits are closely related to  
``deep MERA'' circuits \cite{kim2017robust,sewell2022thermal} but are composed of Gaussian matchgates Eq.~(\ref{eq:matchgate}).
In contrast to the GMPS circuit which constructs a single layer of gates in one long pass over the system, the GMERA circuit makes a logarithmic number of passes over the system as shown in Fig.~\ref{fig:GMERAdemo}(a). 
In this way, higher layers of the circuit introduce longer-range basis transformations in real space, capturing longer-range correlations. 

Here we are primarily interested in how the entanglement-based renormalization group carried out by the GMERA coarse grains the impurity problem and gives rise to an emergent 
effective Hamiltonian with fewer degrees of freedom, in one case resembling a Wilson chain from the numerical renormalization group approach.
Similar to the GMPS construction, we only apply the GMERA circuit to the bath electrons as shown in Fig. \ref{fig:GMERAdemo}(a) to avoid 
spreading the interaction terms over multiple sites in the transformed basis.

In each layer/iteration, a GMERA circuit starts from the bath sites closest to the impurity, whose real space location scales exponentially as $r=2,3,5,\cdots$ (or $r=2^k+1$ with $k=0,1,2,\cdots$). 
The entanglement renormalization generates effective \emph{bath hybridizations} directly from the impurity to these logarithmic number of sites, reflected by the first row and column in the GMERA-transformed bath in Fig.~\ref{fig:GMERAdemo}(b). 

The most significant hopping terms in the  transformed bath form a pattern resembling the so-called ``star geometry'' form of an impurity problem \cite{PhysRevB.90.235131,bauernfeind2019comparison}
\begin{equation}\label{eq:star}
\begin{split}
    H^{\mathrm{star}}_{\mathrm{eff}}\equiv H_{\mathrm{imp}}+ \sum_{k\in \mathcal{A}, \sigma} \bigg[\tilde{V}_{k} (d^\dagger_\sigma c_{k\sigma}+ \mathrm{H.c.}) + \tilde{\varepsilon}_k c_{k\sigma}^\dagger c_{k\sigma}\bigg],
\end{split}
\end{equation}
where there are no hoppings between bath sites and only hybridizations $V_k$ to the impurity site.
Here $k$ denotes the GMERA orbitals that run over the set $\mathcal{A}$ of coarse-grained active orbitals that are strongly entangled with the impurity.  The matrix elements $\tilde{\varepsilon}_k, \tilde{V}_{k}$ are the transformed bath couplings to the impurity, that is, the diagonal and the first row of 
Fig. \ref{fig:GMERAdemo}(b). 

Since the entanglement renormalization procedure does not project out diagonal components of the Hamiltonian, the inactive (diagonal) set of orbitals, $\mathcal{I}$, needs to be included explicitly as 
\begin{equation}
\begin{split}
        H^{\mathrm{GMERA}}_{\mathrm{eff}}= H^{\mathrm{star}}_{\mathrm{eff}} +\sum_{k\in \mathcal{I},\sigma} \tilde{\varepsilon}_k c_{k\sigma}^\dagger c_{k\sigma},
\end{split}
\end{equation}
for the ground state energy to be preserved.

The decoupling of the complementary sites from the impurity can be further verified by computing the mutual information matrix, which is defined as
\begin{equation}
\begin{split}
I(r_1, r_2) &= S(\rho_{\{r_1\}})+S(\rho_{\{r_2\}})-S(\rho_{\{r_1,r_2\}}),
\end{split}
\end{equation}
where $\rho_{\{\cdots\}}$ denotes the reduced density matrix for the sites specified inside the curly brackets. 
In Fig. \ref{fig:GMERAdemo}(c), numerical results including a nonzero interaction $U$ indicate that the mutual information is strongly peaked at these exponential sites while the majority of the remaining sites share negligible mutual information with the impurity 
\footnote{As a technical but important detail, in order to completely  diagonalize the correlation matrix for a finite-sized system, the right boundary of the chain is treated by the GMPS procedure, which creates large non-zero mutual information at the tail as well as some finite values at limited intermediate sites.}. 

\begin{figure}[t!]
\begin{center}
\setlabel{pos=nw,fontsize=\large,labelbox=false}
\includegraphics[width = 0.46\textwidth]{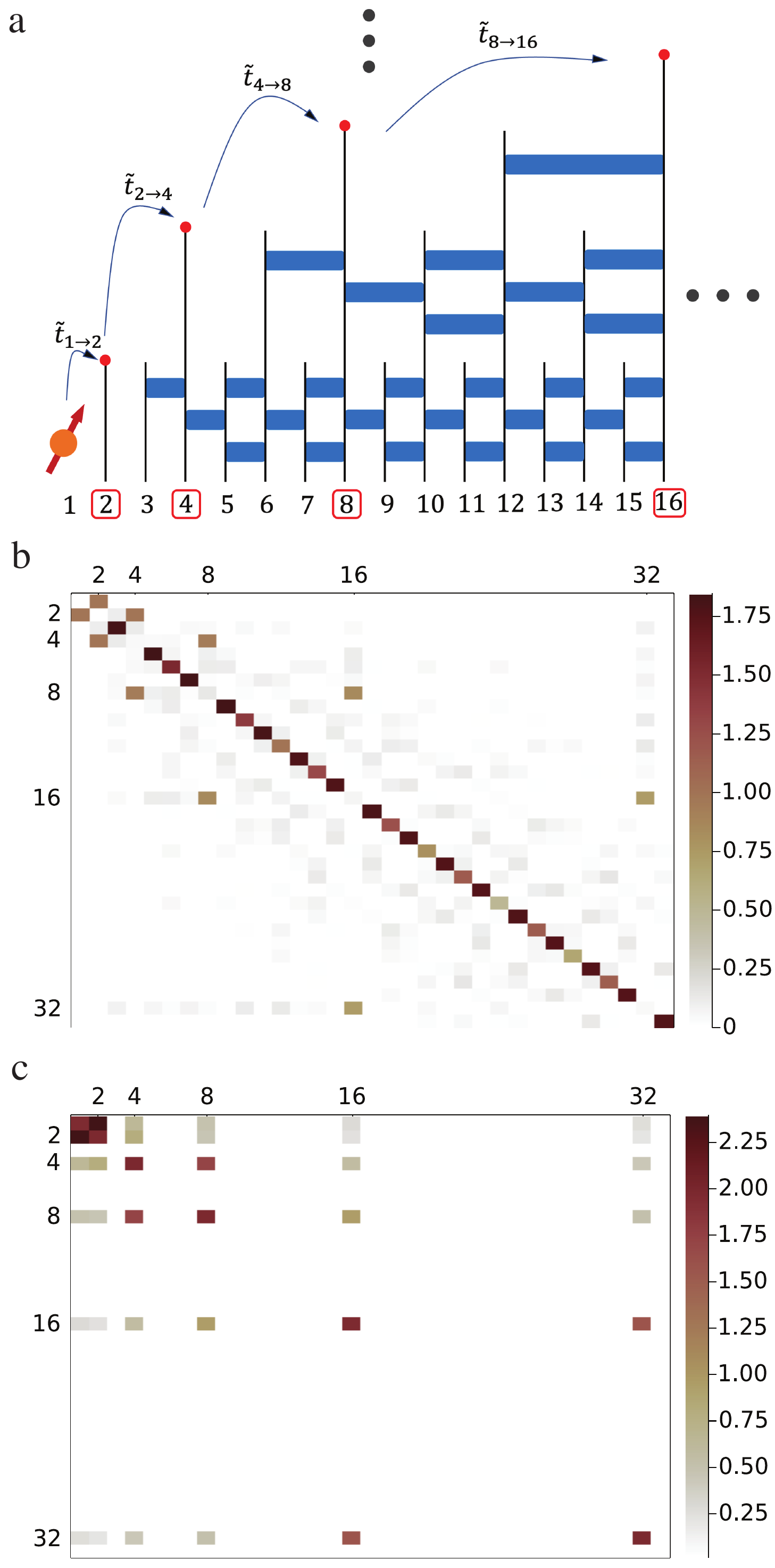}
\caption{\label{fig:GMERA2demo} Demonstration of a boundary GMERA circuit. (a) The boundary GMERA circuit obtained by diagonalizing even sites in each iteration with system size $N=16$ and block size $B=4$. The impurity acts as a local spin in the left. The boundary GMERA circuit generates long-range hoppings among exponential sites, $r=2^k, k=0, 1,\cdots$, marked by the red dots and red squares. (b) The transformed bath Hamiltonian under the boundary GMERA circuit from bath electrons with $N=80, V=t$, where the effective hoppings (color bar) are prominent. (c) The part of mutual information matrix $I(r_1, r_2)$ of the ground state with $U=t$ for $N=80, V=t$. The peaks in the mutual information (color bar) perfectly fit with exponential sites appeared in panels (a) and (b).
}
\end{center}
\end{figure}

An interesting variation of the GMERA circuit is to omit the part of each layer closest to the impurity, as shown in
 Fig. \ref{fig:GMERA2demo}(a). We call this pattern the boundary GMERA (bGMERA) circuit. This is reminiscent of the boundary MERA introduced in Refs. \onlinecite{PhysRevB.82.161107,evenbly2014algorithms,PhysRevB.91.205119}, a modification of traditional MERA tensor networks in the presence of boundaries and impurities. 
Instead of fully diagonalizing the correlation matrix, this procedure intentionally keeps the first site in each iteration untreated. 
These untreated sites also scale exponentially in distance in the transformed basis, $r=2^k,$ with $k=1, 2,\cdots$ and create long-range hopping in the transformed bath Hamiltonian as shown in Fig. \ref{fig:GMERA2demo}(b). The pattern from the transformed Hamiltonian provides a coarse-grained version of the chain-like geometry of the original model [Eq.~\ref{eq:H_A}, $H^{\mathrm{chain}} = H_\mathrm{A}$]
\begin{equation}\label{eq:chain}
\begin{split}
        H^{\mathrm{chain}}_{\mathrm{eff}}&= H_{\mathrm{imp}}+\sum_\sigma \bigg[\tilde{V} (d^\dagger_\sigma c_{2\sigma}+ \mathrm{H.c.})+\tilde{\varepsilon}_2 c_{2\sigma}^\dagger c_{2\sigma}\bigg]\\
        &+ \sum_{k\in \mathcal{A}, \sigma} \bigg[\tilde{t}_{k} (c^\dagger_{k\sigma} c_{k+1\sigma}+ \mathrm{H.c.}) +\tilde{\varepsilon}_{k+1} c_{k+1\sigma}^\dagger c_{k+1\sigma}\bigg],
\end{split}
\end{equation}
where $k$ is the boundary GMERA orbitals and the set $\mathcal{A}$ is the set of coarse-grained active orbitals from boundary GMERA. The transformed hybridization $\tilde{V}$  and matrix elements $\tilde{\varepsilon}_k,\tilde{t}_{k}$ can be read out from the transformed quadratic part of the Hamiltonian, i.e., the diagonal and the prominent hopping at sites $2^k$ of Fig. \ref{fig:GMERA2demo}(b). Again, for the energy to match the original, untransformed system, the inactive complementary set of orbitals, $\mathcal{I}$ has a purely diagonal Hamiltonian which we can add to form the total Hamiltonian as
\begin{equation}
\begin{split}
        &H^{\mathrm{bGMERA}}_{\mathrm{eff}}= H^{\mathrm{chain}}_{\mathrm{eff}} +\sum_{k\in \mathcal{I},\sigma} \tilde{\varepsilon}_k c_{k\sigma}^\dagger c_{k\sigma} \ .
\end{split}
\end{equation}
The mutual information, shown in Fig. \ref{fig:GMERA2demo}(c), elucidates the pattern of correlations between the active and inactive orbitals and the impurity.
This is consistent with previous results on boundary MERA where a Wilson-chain-like structure \cite{PhysRevB.82.144409} naturally arises from the MERA tensor network in the presence of impurities \cite{PhysRevB.82.161107,evenbly2014algorithms,PhysRevB.91.205119}. Here we also find the emergent long-range hopping terms [significant off-diagonal entries in Fig.~\ref{fig:GMERA2demo}(b)] also form a chain of hoppings between the active sites, with values decaying with distance from the impurity. We plan to investigate the connection to the Wilson chain construction in more depth in a future work.

\begin{figure}[t!]
\begin{center}
\includegraphics[width = 0.49\textwidth]{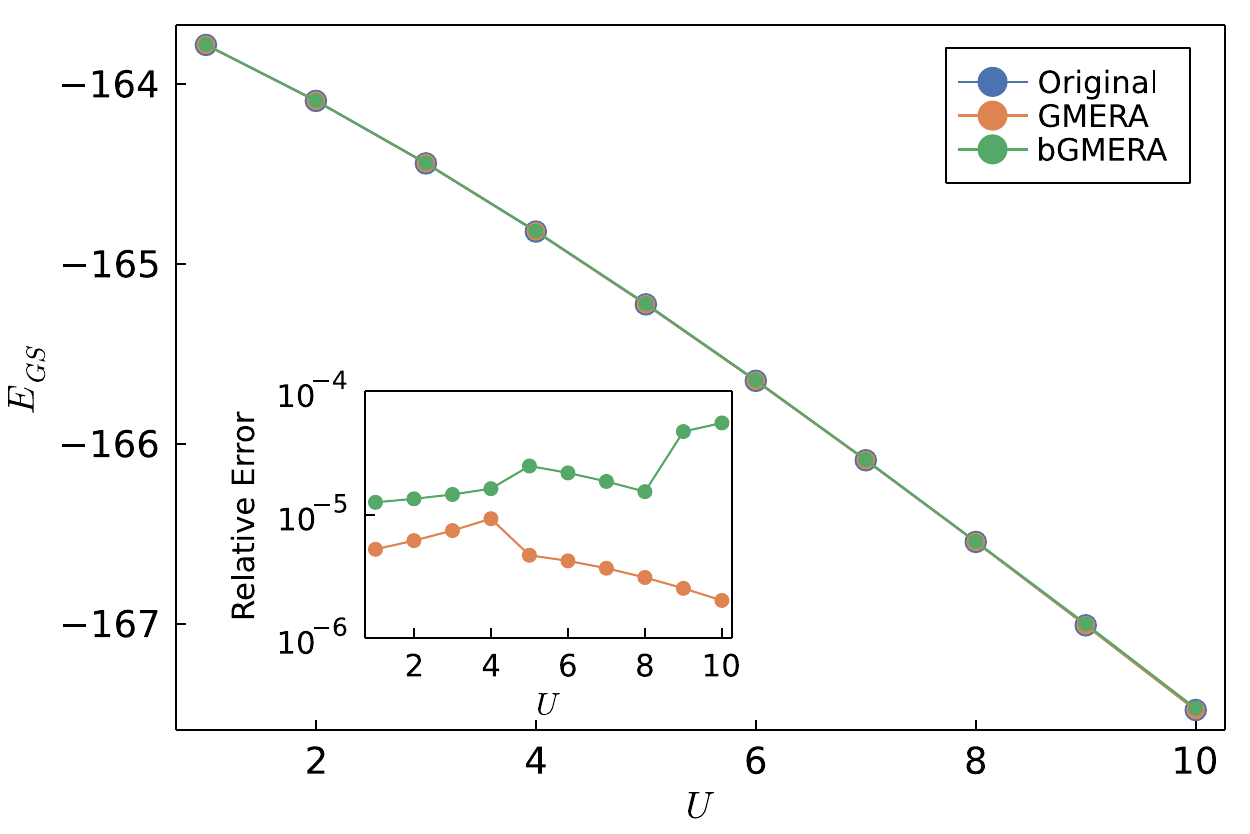}
\caption{\label{fig:EGSscale} The ground state energy as a function of interaction strength $U$ with $N=129, V=t, \epsilon_d=-U/2$. The GMERA energies are obtained by finding the ground state in the emergent Hamiltonian,$H^{\mathrm{chain}}_{\mathrm{eff}},H^{\mathrm{star}}_{\mathrm{eff}}$ , plus single-particle energies smaller than the chemical potential, $\mu=0$. The relevant sites are $N_\mathrm{GMERA}=22,N_\mathrm{bGMERA}=14$, respectively, which are selected from the mutual information with cutoff $I(1,r)\ge 10^{-3}$.
Inset: the relative error of the emergent Hamiltonian benchmarked by the ground state in the original basis.
}
\end{center}
\end{figure}

The correctness of the above two emergent reduced effective
Hamiltonians for approximating the ground-state energy is surprisingly good, as shown in Fig. \ref{fig:EGSscale}. For a system of size $N=129$, a reduced system of \mbox{$N_\mathrm{bGMERA}=14$} sites can already accurately capture the ground state energy for a wide range of interactions $U/t \in [1, 10]$ with an error less than $0.1\%$ \footnote{Since we focus on the interaction at the impurity site $1$ and the impurity mutual information (first row of the matrix) contains all necessary entanglement information, emergent models are taken from the impurity mutual information $I(1,r)$ with cutoffs, for both ground state energies and dynamics.}.
Note that such great agreement of ground-state energies is not obvious purely from the transformed bath Hamiltonian since some neglected terms of the Hamiltonian have coefficients with absolute values greater than $0.1$. Thus, entanglement renormalization of the GMERA circuits yields nontrivial emergent models for the interacting problem. Moreover, the emergent models also provide efficiently coarsed-grained models for time evolution. In Appendix \ref{app:GMERATevo}, the lesser Green's function obtained in small emergent models ($<50$ sites) can accurately capture the behavior in the originally large system ($256$ sites).

\section{Conclusion and Discussion}
\label{sec:conclusion}

Fermionic Gaussian circuits obtained from noninteracting states can drastically reduce the entanglement of impurity-model systems. Circuits known as Gaussian matrix product states (GMPSs) are able to efficiently disentangle many-body ground states across a wide range of interactions realizing different fixed-point physics of the impurity problem. Furthermore,  transforming the Hamiltonian before performing a DMRG calculation can yield much faster DMRG runtimes and more accurate properties for a given truncation threshold.

We studied the Kondo screening cloud under the GMPS circuit transformation, finding that the entanglement remaining after a GMPS transformation reveals universal information about Kondo screening physics. Moreover, when applying the GMPS circuit to two-impurity models, the transformed basis reveals correlations and quantum entanglement between the two impurities that is mediated through the conduction electrons, which is a  manifestation of the RKKY interaction.   We also studied the potential for GMPS transformations to reduce the entanglement of systems undergoing dynamics, where the entanglement growth of a quenched, noninteracting system was found to hardly increase with system size after the transformation, in stark contrast to the behavior in the original real space basis.

We then tested a variation of fermionic Gaussian circuits inspired by the multiscale entanglement renormalization ansatz (MERA) tensor network, called the fermionic Gaussian MERA (GMERA). Such a renormalization introduces a hierarchical structure that induces effective long-range interactions. In this renormalized basis an effective reduced system is identified, which is comprised of a logarithmic number of sites that are highly entangled to the impurity (or each other, depending on the construction), from which a smaller effective Hamiltonian emerges and the rest of the sites can be treated as decoupled free electron modes. Such a bold approximation works exceptionally well in finding the ground state energies for a wide range of interactions and can also efficiently simulate the impurity Green's function dynamics, over a time range subjected to finite size effect. The essential functionalities for conveniently working with fermionic Gaussian circuits are now available in the Julia ITensor ecosystem \footnote{\url{https://github.com/ITensor/ITensors.jl/tree/main/ITensorGaussianMPS}}.

An important future consideration is the relationship between the orbitals we use to transform the basis and the natural orbitals, which are the exact eigenvectors of the correlation matrix that are generally nonlocal.
The natural orbital basis has been used widely in quantum chemistry applications and in approaches for solving impurity problems \cite{PhysRevB.90.085102,PhysRevB.86.165128,PhysRevB.88.035123,PhysRevB.89.085108,PhysRevB.100.115134}.
In Ref.~\onlinecite{Krumnow:2021}, it was observed that the mode transformations optimized to minimize entanglement (different from ours) sometimes coincided with the natural orbitals. The orbitals we find come from a process which balances locality, entanglement reduction, interpretability, and diagonalization of the correlation matrix whereas other works consider only one or two of these goals. In particular, we compare our basis to the single-particle rotation used in Ref.~\onlinecite{Kohn:2021} for computing the impurity Green's function in the interacting Anderson impurity model.



There are many more directions to explore in applying fermionic Gaussian circuit transformations to interacting systems. 
Based on the results from time evolution, we believe GMPS circuits could also keep entanglement of the interacting problem under control.
When using the GMERA circuit, it will be interesting to see how well the emergent sites from noninteracting quenched states still work in time evolution when including interactions.
In a separate but related vein, recent work has demonstrated how to efficiently convert the so-called influence matrix in time evolution of interacting impurity models \cite{Thoenniss}, which shows a potential usage of fermionic Gaussian circuits in nonequilibrium quantum dynamics under noninteracting fermionic reservoirs.

More challenging directions involve disentangling systems with spatially uniform interactions, such as the one-dimensional Hubbard model.
In general, we believe the application of fermionic Gaussian circuits could push tensor network methods and DMRG toward handling much more challenging and highly entangled problems with a proper choice of the fermionic Gaussian circuit transformation, potentially even in higher dimensions, i.e.~general tree tensor networks, MERA circuits in 2D systems, and Gaussian projected entangled pair states for 2D fermionic problems. 



\section*{Acknowledgments}
We thank Xiaodong Cao, Chia-Min Chung, Jan von Delft, \"Ors Legeza, Olivier Parcollet, and Andreas Weichselbaum for constructive discussions. 
All calculations were performed with ITensor software \cite{fishman2022itensor,itensor-r0.3} in the Julia language \cite{Bezanson_Julia_A_fresh_2017}.
A.-K. W. and J.H.P. are partially supported by the National Science Foundation (NSF)
CAREER Grant No. DMR-1941569 and the Alfred P. Sloan Foundation through a Sloan Research Fellowship. 

\appendix

\section{Split-site Representation}
\label{app:splitsite}
In Sec.~\ref{sec:model} we introduced the split spin representation of the Anderson impurity model, here in this appendix we describe this in more detail.
Without changing the physics, the impurity model can be rewritten by separating spin up and down operators $a_{-j}\equiv c_{j\uparrow}, a_{j}\equiv c_{j\downarrow}$, which has been used often for studying impurity systems with MPS techniques \cite{Saberi:2008,Ganahl:2015,Rams:2020,Kohn:2021}.

Specifically, the model can be rewritten in terms of spinless fermion operators by putting all up spins on the left side of the impurity and all down spins on the right side. The impurity becomes two spinless fermion sites in the center of the system. Quantitatively:
\begin{equation}
\begin{split}
    H_\mathrm{A}&=H_{\uparrow}+H_{\mathrm{imp}}+H_\downarrow,
\end{split}
\end{equation}
with
\begin{equation}\label{eq:model}
\begin{split}
    &H_{\uparrow}=-t\sum_{j=-2}^{-(N-1)} (a_{j}^\dag a_{j-1}+\text{H.c.}) -V(a_{-1}^\dag a_{-2}+\text{H.c.})\\
    &H_{\mathrm{imp}}=\epsilon_d (\hat{n}_{-1}+\hat{n}_{1}) + U\hat{n}_{-1}\hat{n}_{1}\\
    &H_\downarrow=-V(a_{1}^\dag a_{2}+\text{H.c.})-t\sum_{j=2}^{N-1}(a_{j}^\dag a_{j+1}+\text{H.c.}).
\end{split}
\end{equation}
 Here $a_j$ is a spinless fermonic operator. $j<0$ represents original site $r=-j$ for up spins while $j>0$ is site $r=j$ for down spins, and $-1, 1$ sites are the impurity site for up and down impurity spins (index $0$ is skipped). $\hat{n}_j=a^\dag_j a_j$ is the occupancy operator. For all cases, we keep the particle-hole symmetry for the impurity $U=-2\epsilon_d$. 

As alluded to earlier, the primary advantage of this model geometry is that the ground state bond dimension $\chi$ is decreased by approximately square root of the original bond dimension (which becomes exact in the noninteracting limit $U=0$) for the same MPS accuracy. The trade-off is that computing operators such as the spin-spin correlators $C(r) = \langle \mathbf{S}(1)\cdot \mathbf{S}(r)\rangle$ and the bipartite impurity entanglement entropy $S(r)$ in real space become slightly more complicated to compute.


\section{Entanglement entropy of a middle segment in an MPS (ABA partition)}
\label{app:impurity-entropy}

To obtain the entanglement entropy of a cloud of radius $r$ around the impurity in real space, we need to divide the split-site geometry of the 1D chain into three partitions, where the entanglement entropy of interest is between the middle partition of sites $j\in [-r, r]$ and the rest of the system. The orthogonality or the canonical form of the MPS is crucial in obtaining the entanglement of this ABA partition. To begin with, it is helpful to revisit how the bipartite entanglement is computed. When we decompose a 1D chain into blocks of $A$ and $B$, where block $A$ goes from $1$ to $l$ while $B$ goes from $l+1$ to $N$, we can order the MPS in the following way
\begin{equation}
    \begin{split}
    \ket{\Psi_0}&=\sum_{\sigma_1,\cdots,\sigma_{2N}} c_{\sigma_1,\cdots,\sigma_{2N}}\ket{\sigma_1,\cdots,\sigma_{2N}}\\
        &=\sum_{\sigma_1,\cdots,\sigma_{2N}} M^{\sigma_1}M^{\sigma_2}\cdots M^{\sigma_{2N}}\ket{\sigma_1,\cdots,\sigma_{2N}}\\
        &=\sum_{a_l} \ket{a_l}_A\ket{a_l}_B,
    \end{split}
    \label{eqn:cs}
\end{equation}
where 
\begin{equation}
    \begin{split}
        \ket{a_l}_A &=\sum_{\sigma_1,\cdots,\sigma_{l}} (M^{\sigma_1}\cdots M^{\sigma_l})_{1,a_l} \ket{\sigma_1,\cdots,\sigma_{l}},\\
        \ket{a_l}_B &=\sum_{\sigma_{l+1},\cdots,\sigma_{N}} (M^{\sigma_{l+1}}\cdots M^{\sigma_N})_{a_l,1} \ket{\sigma_{l+1},\cdots,\sigma_{N}}.
    \end{split}
\end{equation}
The boundary tensors $M^{\sigma_1}, M^{\sigma_N}$ give $1$ in the subscript. Note that this decomposition $\ket{\Psi_0}=\sum_{a_l} \ket{a_l}_A\ket{a_l}_B$ is not a Schmidt decomposition, because, in the left-canonical form of $M^\sigma$, the states $\ket{a_l}_A$ form an orthonormal set for system $A$ while states $\ket{a_l}_B$ do not. Thus, to perform the Schmidt decomposition, we apply the singluar value decomposition (SVD) to the index/bond between $M^{\sigma_{l}}$ and $M^{\sigma_{l+1}}$, where the diagonal $S$ matrix gives the singular values.

\begin{figure}[t!]
\begin{center}
\includegraphics[width = 0.48\textwidth]{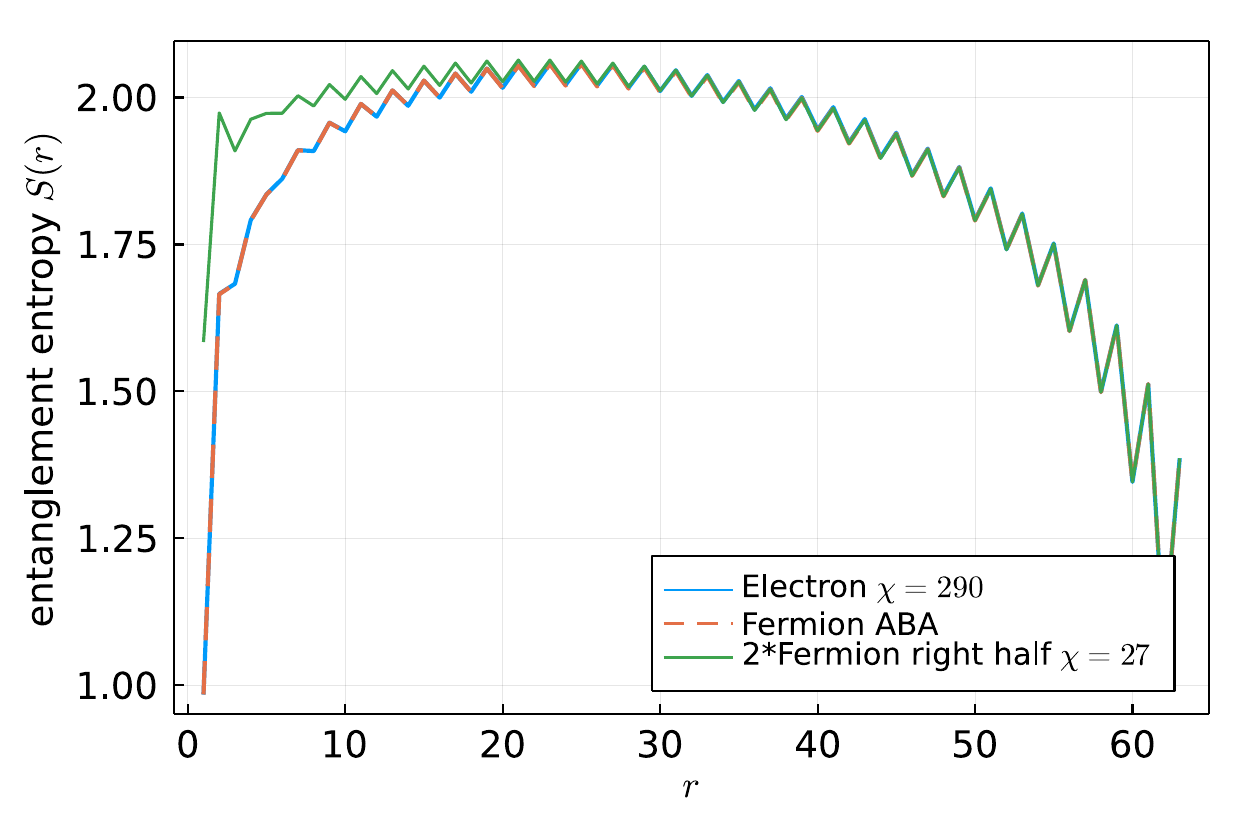}
\caption{\label{appfig:entropy_compare} Entanglement entropy in different geometries of the single impurity Anderson model with $N=64, V=1.0t, U=8t, \epsilon_d=-U/2$. Solid blue curve represents bipartite entropy in the electron name space of the Hamiltonian where the impurity locates in the left. Red dashed curve is the entanglement entropy of a middle segment of the split-site MPS with a radius $r$ around the impurity sites. Green curve is the double of the right half of the bipartite entropy of the split-site MPS, demonstrating that it is quantitatively similar to the electron entanglement entropy sufficiently far from the impurity.}
\end{center}
\end{figure}

For the ABA partition, where the middle part runs from $l_1+1$ to $l_2$, we want to use a mixed-canonical MPS automatically facilitated by ITensor, 
which allows us to represent the wavefunction amplitudes in Eq.~\eqref{eqn:cs} as
\begin{equation}
    \begin{split}
        c_{\sigma_1,\cdots,\sigma_{N}}&=\sum_{\sigma_1,\cdots,\sigma_{N}} L^{\sigma_1}\cdots L^{\sigma_{l_1}} S R^{\sigma_{l_1+1}}\dots R^{\sigma_{N}},\\
    \end{split}
\end{equation}
where $L^\sigma, R^\sigma$ represent left and right normalized, $\sum_\sigma (L^\sigma)^\dagger L^\sigma=\sum_\sigma R^\sigma (R^\sigma)^\dagger = \mathbb{I}$ respectively with the orthogonality center at site $l_1+1$. The decomposition we look for is
\begin{equation}
    \begin{split}
        \ket{\Psi_0}&=\sum_{b} \ket{a}_A\ket{b}_B \ket{c}_C,\\
        \ket{a}_A &= \sum_{\sigma_1,\cdots,\sigma_{l_1}} (L^{\sigma_1}\cdots L^{\sigma_{l_1}})_{1,a} \ket{\sigma_1,\cdots,\sigma_{l_1}},\\
        \ket{b}_B &= \sum_{\sigma_{l_1+1},\cdots,\sigma_{l_2}} (SR^{\sigma_{l_1+1}}\cdots R^{\sigma_{l_2}})_{a,c} \ket{\sigma_{l_1+1},\cdots,\sigma_{l_2}},\\
        \ket{c}_C &= \sum_{\sigma_{l_2+1},\cdots,\sigma_{N}} (R^{\sigma_{l_2+1}}\cdots R^{\sigma_{N}})_{c,1} \ket{\sigma_{l_2+1},\cdots,\sigma_{N}},
    \end{split}
\end{equation}
where for both subsystem $A$ and $C$, the states are orthogonal and states $\ket{b}_B$ will have a size of $a\times c$. However, if we directly extract the middle segment, then we will still work on a huge tensor with physical space $2^{l_2-l_1}$ for the SVD while most information within the middle segment is redundant. Therefore, we can instead contract these physical indexes $l_1+1,\cdots,l_2$ but only leave the $2$ bond indexes by multiplying with complex conjugate of $\ket{b}_B$ as the compressed density matrix. In the end, the SVD will work on a reduced tensor, consisting of $4$ indices $a, a', c, c'$, by 2 indices $a, c$. Graphic instructions for this calculation appear in the Appendix A of Ref. \onlinecite{rogerson2022entanglement}.

When comparing the split-site representation of the Hamiltonian with the original electron representation as in Fig. \ref{appfig:entropy_compare}, it is clear that the bipartite entanglement entropy in the electron space is the same as the middle segment entropy of the split-site MPS. 
Sufficiently far away from the impurity, the entanglement entropy is well described by twice that of the split-site fermion MPS. 

\section{Variations of the GMPS circuits}
\label{app:diffGMPS}

\begin{figure}[t!]
\begin{center}
\includegraphics[width = 0.49\textwidth]{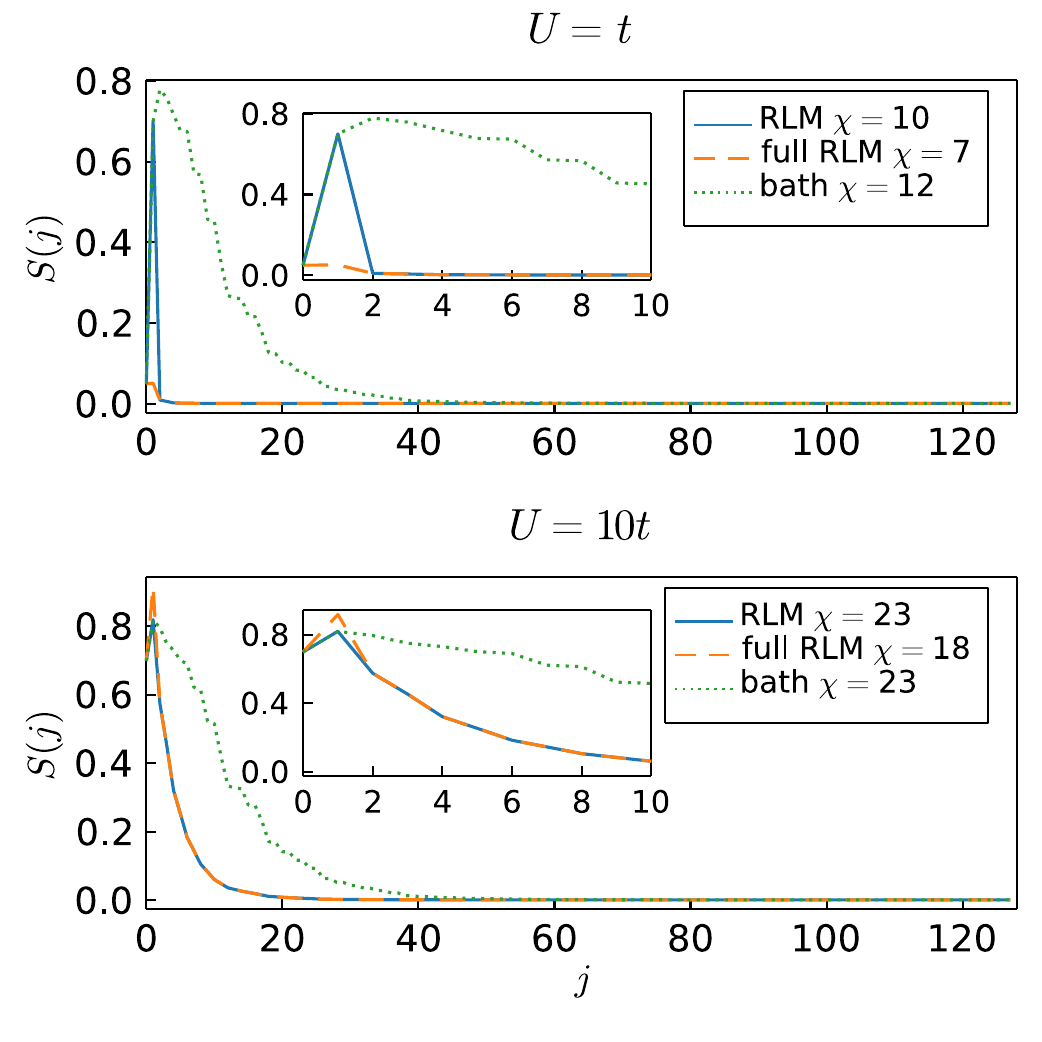}
\caption{\label{fig:vargates}  Disentangling effect of different GMPS circuits for system $N=128, V=t$ for the small repulsion $U=t$ (upper panel) and the large repulsion $U=10t$ (lower panel) cases. The solid curves are the same results as the main text disentangled by the GMPS circuit from the resonant level model with impurity gates dropped in the end. The orange dashed curves are results under the whole GMPS circuit without dropping gates. The green dotted curves are results under the GMPS circuit obtained only from bath electrons. 
}
\end{center}
\end{figure}

In the main text, the GMPS circuits are obtained from the $U=0$ limit of the impurity model (known as the resonant level model, RLM) with impurity gates dropped to maintain locality of the Coulomb repulsion $U$. Such a treatment benefits the computation without changing the qualitative disentangling behavior. As shown in Fig.~\ref{fig:vargates}, the disentangled results are very similar between the circuits with and without gates involving the impurity. Though, the full circuit gives less entanglement at the impurity site in the small $U$ limit, the entanglements away from the impurity are basically the same with the circuit that avoids the impurity, which becomes more prominent in large $U$ limit.

The GMPS circuit obtained from the bath electrons is also considered in Fig.~\ref{fig:vargates}. Compared to the resonant-level model, disentangling results of the circuit from the bath has higher entanglement over a large range of sites around the impurity. Qualitatively, we attribute this to the electron MPS having to represent the impurity bound state as a superposition over a large number of bath sites, which is already taken into account exactly in the resonant-level model MPS. 

\section{Hamiltonian MPO with hopping cutoff}
\label{app:MPOcutoff}

\begin{figure}[b!]
\begin{center}
\includegraphics[width = 0.48\textwidth]{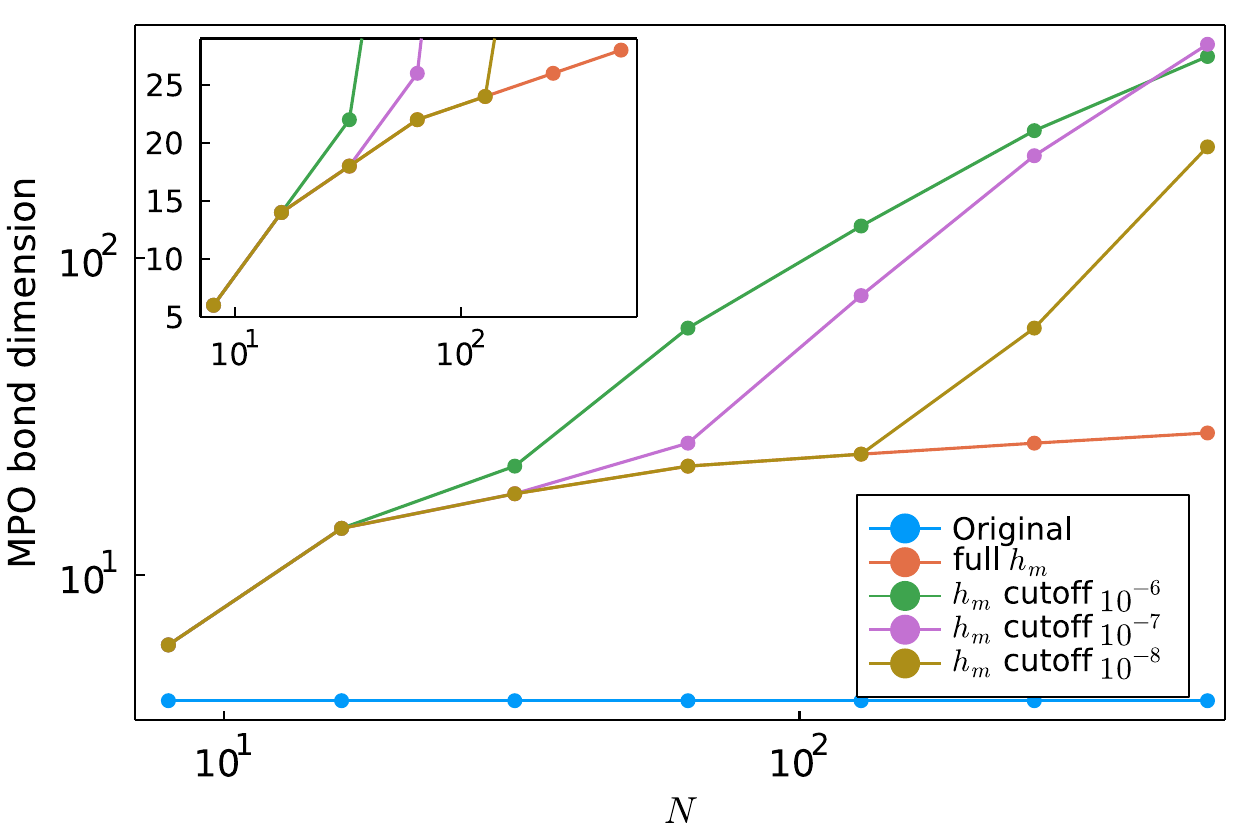}
\caption{\label{fig:MPOcutoff} The maximum bond dimensions $\chi$ as a function of physical sizes $N$ with various hopping cutoffs with $V=t, U=t$. Given a finite cutoff, the maximum MPO bond $\chi$ dramatically increases at a large system size. Inset: same results in log-$x$ scale.
}
\end{center}
\end{figure}

In Sec. \ref{sec:MPO} of the main text, the GMPS circuits are used to transform the Hamiltonian, which generate quadratic number of terms. To constrain the number of nonzero hopping terms $t_{kl}$ in the transformed noninteracting Hamiltonian, $\tilde{H}_\uparrow=R^\dagger H_\uparrow R=\sum_{kl}t_{kl}f_k^\dagger f_l$, we tried imposing a hopping size cutoff, whose effects on the resulting compressed MPO are studied here. 
 As shown in Fig.~\ref{fig:MPOcutoff}, the maximum MPO bond dimensions does not change with system sizes in the original basis, while it increases logarithmically in the system size $N$ asymptotically when we use all nonzero terms in the transformed basis.
 Perhaps counterintuitively, attempting to truncate small hopping terms according to a fixed cutoff leads to MPOs with larger bond dimensions.
 Thus, small hopping terms have nontrivial contribution in compressing the Hamiltonian into an MPO. This effect can possibly be understood as the hoppings having a smooth decaying behavior, so that truncating them introduces artificial ``steps'' that can make the MPO compression algorithm less effective. Therefore in practice we keep all transformed Hamiltonian terms.

\section{Speedup in two-impurity model} 
\label{app:twoImpSpeedup}

\begin{figure}[b!]
\begin{center}
\includegraphics[width = 0.49\textwidth]{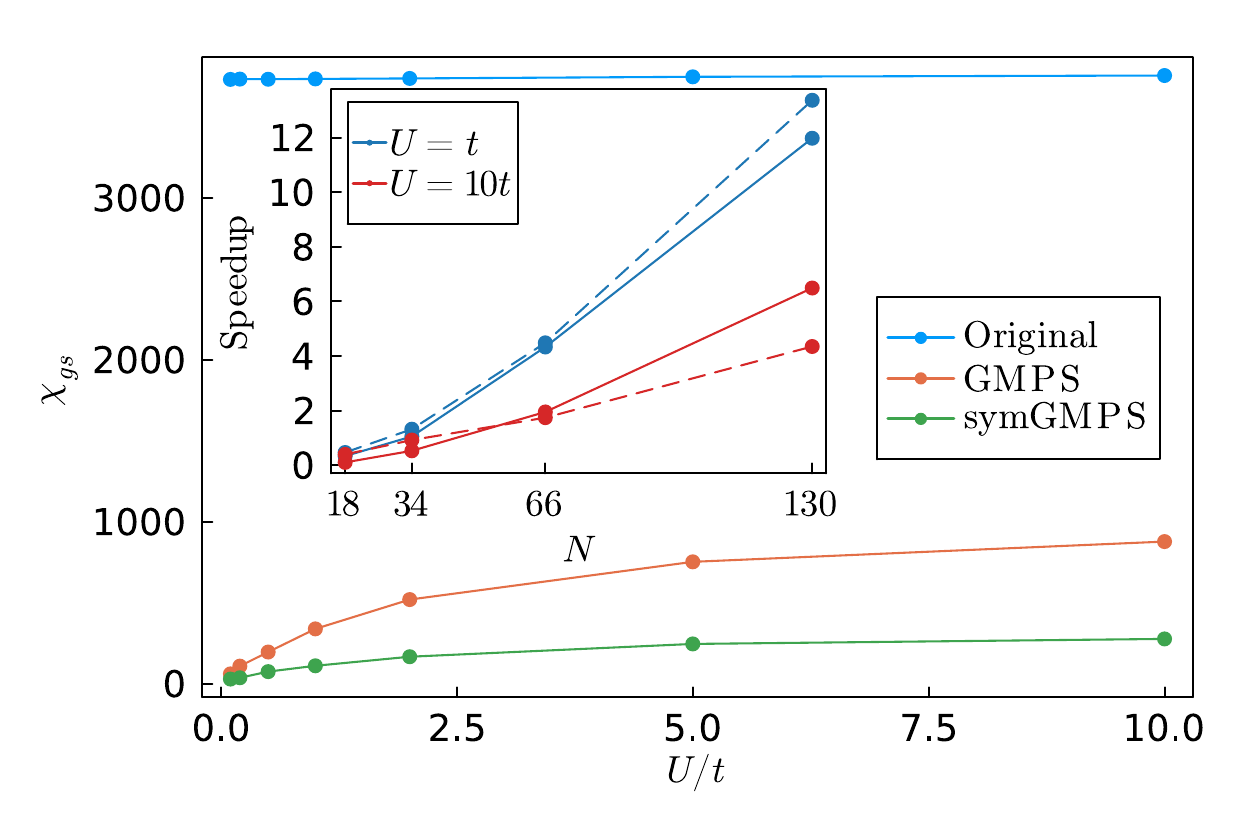}
\caption{\label{fig:Bs2Iimp} The maximum MPS bond dimensions as a function of interaction $U$ for the two-side geometry with $N=2^7+2$ and $V=t$. Inset: the speedup of the median DMRG sweep time. Solid curves are in the symGMPS basis while dashed curves are in the GMS basis. 
}
\end{center}
\end{figure}

As fermionic Gaussian circuits will inevitably make the interaction nonlocal on the symGMPS basis, the computational cost to obtain the ground states of two-impurity systems depends on the strength of the interaction $U$. Empirically, the side geometry gives the most prominent disentangling effect. As shown in Fig. \ref{fig:Bs2Iimp}, the MPS bond dimension gradually increases with $U$ in disentangled bases, while the overall speedup still increases with larger system sizes.

\section{Time-doubling trick for Green's function evaluation}
\label{app:timedoubling}

We can make use of the fact that the zero-temperature single-particle Green's function is evaluated on the ground state of the system, $\ket{\Psi_0}$ with energy $E_0$ to obtain the correlator at time $t$ from time evolution up to time $t/2$.
\begin{equation}
\begin{split}
&\bra{\Psi_0} c^\dagger(t) c(0) \ket{\Psi_0} = e^{-iE_0t} \bra{\Psi_0} c^\dagger e^{-iHt}  c \ket{\Psi_0} \\
&=e^{-iE_0t} (\bra{\Psi_0} c^\dagger e^{-iHt/2}) (e^{-iHt/2} c \ket{\Psi_0})
\end{split}
\end{equation}
Note that convergence of tensor network state calculations needs to be assessed separately for Green's functions computed according to the first or second line of this equation. Convergence within some tolerance up to a time $t^*$ for Green's functions computed with full time-evolution does not guarantee convergence up to time $2t^*$ for those computed using the time-doubling trick. However, we can generally assume that the convergence time for the latter will be larger than for the former.

\section{Time Evolution under the GMERA Gates } 
\label{app:GMERATevo}

\begin{figure}[b!]
\begin{center}
\includegraphics[width = 0.48\textwidth]{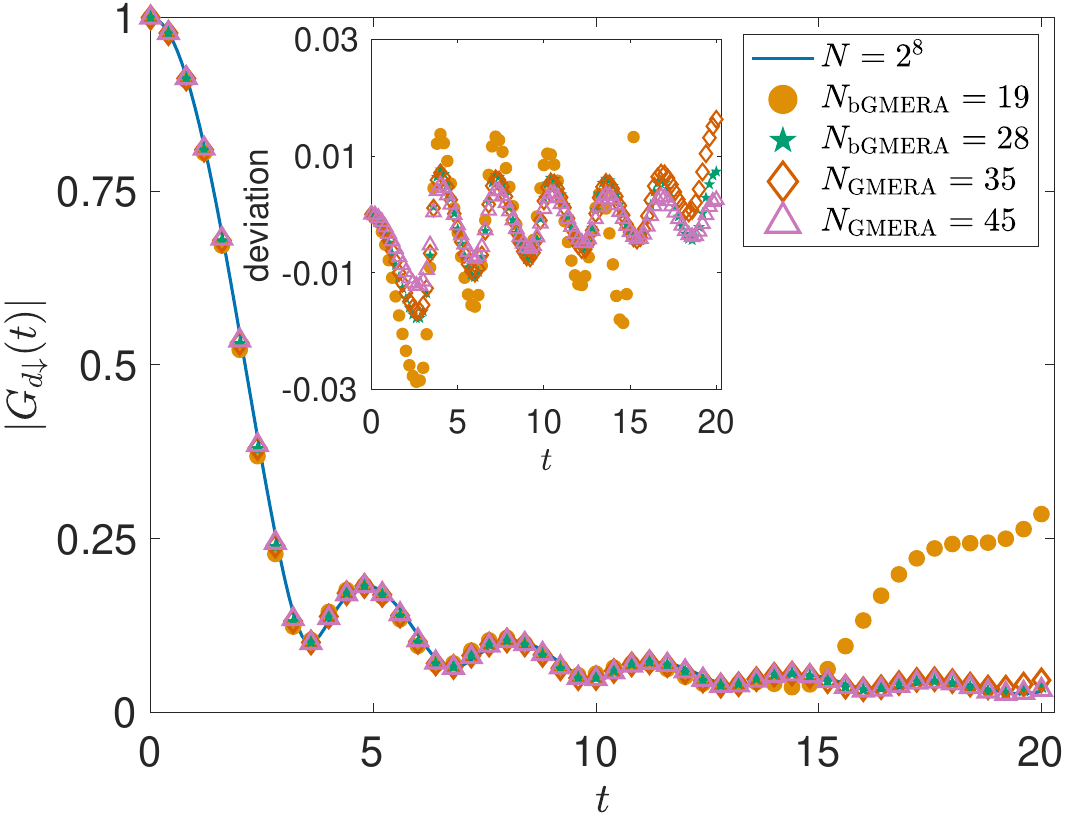}
\caption{\label{fig:Gt} The dynamics of the impurity Green's function [Eq. \ref{eq:Gt}] for emergent Hamiltonians with the original system $N=256, V=t$.  The blue solid curve represents Green's function with the full size regardless of basis. The different markers are emergent systems from the transformed basis GMERA (empty markers) and boundary GMERA (bGMERA, solid markers), which is taken from the mutual information of the quenched state $\ket{\Phi(t=10)}$. For each case, the two different sizes of emergent systems (which are labeled in the legend) are obtained at two mutual information cutoffs ($10^{-4}$ and $10^{-5}$). Inset: the deviation of the emergent models away from the full system Green's function.}
\end{center}
\end{figure}

Another application of GMERA circuits is to evolve the quenched state $\ket{\Phi(t)}$ over time from Eq. \ref{eq:quench} and compare the impurity Green's function in the coarse-grained models
\begin{equation}
\begin{split} \label{eq:Gt}
    G_{\downarrow}(t)&\equiv \langle d_\downarrow^\dagger (t) d_\downarrow(0)\rangle \\
    &=e^{iE_0t/\hbar} \bra{\Psi_0}d_\downarrow^\dagger U(t)d_\downarrow\ket{\Psi_0}\\
    &=e^{iE_0t/\hbar} \langle \Phi(0) |\Phi(t)\rangle,
\end{split}
\end{equation}
where $E_0$ is the energy of the ground state. This is an important quantity both for understanding impurity physics and as an input to the dynamical mean-field theory (DMFT) method \cite{RevModPhys.68.13}.
Here, we focus on time evolving noninteracting impurity models.
Previously, the emergent active sites were identified by analyzing the mutual information of ground states. 
Though the ground state shares very similar mutual information peaks with the impurity quenched state at most times \footnote{Note that the mutual information $I(1,r)=0$ for all $r$ immediately after the quench, but then becomes quickly stabilized with fixed peak locations except at certain isolated times.}, we find it is more accurate to obtain emergent sites from the quenched state mutual information.

The dynamics of the impurity Green's function, Eq.~\ref{eq:Gt}, is shown in Fig.~\ref{fig:Gt}. 
Given the same system size $N$, the absolute values of $|G_{\downarrow}(t)|$ are invariant under arbitrary unitary change of basis. In other words, $|G_{\downarrow}(t)|$ stays the same for the GMPS, the GMERA and the boundary GMERA circuits without coarse-graining.
However, if we choose to work only with the active sites $\mathcal{A}$ [from Eqs. \ref{eq:star} and \ref{eq:chain}]
based on the mutual information of $\ket{\Phi(t)}$ at a typical time, e.g., $t=10$, and time evolve the impurity under the resulting emergent Hamiltonian, then the impurity Green's function can be well approximated with an exponentially small system size ($N_\mathrm{GMERA}=45, N_\mathrm{bGMERA}=28$ as compared to the full size $N=256$) for a wide range of time until finite size effects start to dominate. 
Unlike calculating the ground state energies, the inactive sites (sites with occupancies very close to $0$ or $1$) 
only provide a phase change in their eigenstates, which cancel out for the expectation value of the Green's function and thus are irrelevant \footnote{However, they are relevant when determining particle numbers in coarse-grained models. Given the total particle $N-1$ ($-1$ for impurity annihilation), inactive sites give close to $0$ or $1$ diagonal matrix elements in the diagonalized correlation matrix, whose summation $N_\mathcal{I}$ gives the corresponding particle number in the coarse-grained model by $N_{\mathcal{A}}=N-N_\mathcal{I}-1$.}. The fact that the Green's function can be obtained through a logarithmically small system makes GMERA and boundary GMERA circuits a powerful tools for dynamics, though their efficacy is yet to be seen for studying the dynamics of interacting systems.

\bibliography{ref}

\end{document}